\documentclass[twocolumn]{aastex631}

\let\tablenum\relax
\usepackage{siunitx}
\usepackage{booktabs}
\usepackage{ragged2e}
\usepackage{blindtext}
\usepackage{tabularx}
\usepackage{amsmath}
\usepackage{mathtools}
\usepackage{hyperref}
\usepackage{xspace}
\usepackage{xcolor}
\usepackage{placeins}
\usepackage{upgreek}
\usepackage{rotating}
\font\myfont=cmr12 at 11pt

\usepackage{graphicx}
\usepackage{grffile}
\usepackage{hyperref}

\begin{document}

\title{{\myfont THE CHIME/FRB OUTRIGGERS: COMMISSIONING THE HAT CREEK OUTRIGGER AND AN UPDATED CALIBRATION SCHEME FOR MITIGATING RFI}}

\author[0000-0002-5857-4264]{Mattias Lazda}
\affiliation{Dunlap Institute for Astronomy and Astrophysics, 50 St. George Street, University of Toronto, ON M5S 3H4, Canada}
\affiliation{David A. Dunlap Department of Astronomy and Astrophysics, 50 St. George Street, University of Toronto, ON M5S 3H4, Canada}
\email{mattias.lazda@mail.utoronto.ca}

\author[0000-0001-6128-3735]{Nina V. Gusinskaia}
\affiliation{ASTRON, Netherlands Institute for Radio Astronomy, Oude Hoogeveensedijk 4, 7991 PD Dwingeloo, The Netherlands}
\affiliation{Anton Pannekoek Institute for Astronomy, University of Amsterdam, Science Park 904, 1098 XH Amsterdam, The Netherlands}

\author[0009-0005-6633-3945]{Gurman Sachdeva}
\affiliation{Dunlap Institute for Astronomy and Astrophysics, 50 St. George Street, University of Toronto, ON M5S 3H4, Canada}
\affiliation{David A. Dunlap Department of Astronomy and Astrophysics, 50 St. George Street, University of Toronto, ON M5S 3H4, Canada}

\author[0000-0002-3980-815X]{Shion Andrew}
\affiliation{MIT Kavli Institute for Astrophysics and Space Research, Massachusetts Institute of Technology, 77 Massachusetts Ave, Cambridge, MA 02139, USA}
\affiliation{Department of Physics, Massachusetts Institute of Technology, 77 Massachusetts Ave, Cambridge, MA 02139, USA}

\author[0000-0002-0772-9326]{Juan Mena-Parra}
\affiliation{Dunlap Institute for Astronomy and Astrophysics, 50 St. George Street, University of Toronto, ON M5S 3H4, Canada}
\affiliation{David A. Dunlap Department of Astronomy and Astrophysics, 50 St. George Street, University of Toronto, ON M5S 3H4, Canada}

\author[0000-0001-6523-9029]{Mandana Amiri}
\affiliation{Department of Physics and Astronomy, University of British Columbia, 6224 Agricultural Road, Vancouver, BC V6T 1Z1 Canada}

\author[0000-0001-5908-3152]{Bridget C. Andersen}
\affiliation{Department of Astronomy and Astrophysics, University of California Santa Cruz, 1156 High Street, Santa Cruz, CA 95060, USA}

\author[0009-0003-1765-8845]{Alyssa Atkinson}
\affiliation{Dunlap Institute for Astronomy and Astrophysics, 50 St. George Street, University of Toronto, ON M5S 3H4, Canada}
\affiliation{David A. Dunlap Department of Astronomy and Astrophysics, 50 St. George Street, University of Toronto, ON M5S 3H4, Canada}

\author[0000-0003-3772-2798]{Kevin Bandura}
\affiliation{Lane Department of Computer Science and Electrical Engineering, 1220 Evansdale Drive, PO Box 6109, Morgantown, WV 26506, USA}
\affiliation{Center for Gravitational Waves and Cosmology, West Virginia University, Chestnut Ridge Research Building, Morgantown, WV 26505, USA}

\author[0000-0002-8376-1563]{Alice P. Curtin}
\affiliation{Department of Physics, McGill University, 3600 rue University, Montr\'eal, QC H3A 2T8, Canada}
\affiliation{Trottier Space Institute, McGill University, 3550 rue University, Montr\'eal, QC H3A 2A7, Canada}
\affiliation{Anton Pannekoek Institute for Astronomy, University of Amsterdam, Science Park 904, 1098 XH Amsterdam, The Netherlands}

\author[0000-0001-7166-6422]{Matt Dobbs}
\affiliation{Department of Physics, McGill University, 3600 rue University, Montr\'eal, QC H3A 2T8, Canada}
\affiliation{Trottier Space Institute, McGill University, 3550 rue University, Montr\'eal, QC H3A 2A7, Canada}

\author[0009-0003-3736-2080]{Ian Hendricksen}
\affiliation{Department of Physics, McGill University, 3600 rue University, Montr\'eal, QC H3A 2T8, Canada}
\affiliation{Trottier Space Institute, McGill University, 3550 rue University, Montr\'eal, QC H3A 2A7, Canada}

\author[0000-0003-2317-1446]{Jason W.T. Hessels}
\affiliation{Department of Physics, McGill University, 3600 rue University, Montr\'eal, QC H3A 2T8, Canada}
\affiliation{Trottier Space Institute, McGill University, 3550 rue University, Montr\'eal, QC H3A 2A7, Canada}
\affiliation{ASTRON, Netherlands Institute for Radio Astronomy, Oude Hoogeveensedijk 4, 7991 PD Dwingeloo, The Netherlands}
\affiliation{Anton Pannekoek Institute for Astronomy, University of Amsterdam, Science Park 904, 1098 XH Amsterdam, The Netherlands}

\author[0000-0002-5794-2360]{Dant\'e M. Hewitt}
\affiliation{Anton Pannekoek Institute for Astronomy, University of Amsterdam, Science Park 904, 1098 XH Amsterdam, The Netherlands}

\author[0000-0001-9345-0307]{Victoria M. Kaspi}
\affiliation{Department of Physics, McGill University, 3600 rue University, Montr\'eal, QC H3A 2T8, Canada}
\affiliation{Trottier Space Institute, McGill University, 3550 rue University, Montr\'eal, QC H3A 2A7, Canada}

\author[0009-0005-7115-3447]{Kholoud Khairy}
\affiliation{Lane Department of Computer Science and Electrical Engineering, 1220 Evansdale Drive, PO Box 6109, Morgantown, WV 26506, USA}
\affiliation{Center for Gravitational Waves and Cosmology, West Virginia University, Chestnut Ridge Research Building, Morgantown, WV 26505, USA}

\author[0009-0004-4176-0062]{Afrokk Khan}
\affiliation{Department of Physics, McGill University, 3600 rue University, Montr\'eal, QC H3A 2T8, Canada}
\affiliation{Trottier Space Institute, McGill University, 3550 rue University, Montr\'eal, QC H3A 2A7, Canada}

\author[0000-0003-3457-4670]{Ronniy C. Joseph}
\affiliation{S[\&]T Netherlands, Olof Palmestraat 14, 2616 LR Delft, The Netherlands}

\author[0000-0003-2116-3573]{Adam E. Lanman}
\affiliation{MIT Kavli Institute for Astrophysics and Space Research, Massachusetts Institute of Technology, 77 Massachusetts Ave, Cambridge, MA 02139, USA}
\affiliation{Department of Physics, Massachusetts Institute of Technology, 77 Massachusetts Ave, Cambridge, MA 02139, USA}

\author[0000-0002-4209-7408]{Calvin Leung}
\affiliation{Miller Institute for Basic Research, Stanley Hall, Room 206B, Berkeley, CA 94720}
\affiliation{Department of Astronomy, University of California, Berkeley, CA 94720, United States}

\author[0000-0002-4279-6946]{Kiyoshi W. Masui}
\affiliation{MIT Kavli Institute for Astrophysics and Space Research, Massachusetts Institute of Technology, 77 Massachusetts Ave, Cambridge, MA 02139, USA}
\affiliation{Department of Physics, Massachusetts Institute of Technology, 77 Massachusetts Ave, Cambridge, MA 02139, USA}

\author[0000-0002-8912-0732]{Aaron B. Pearlman}
\altaffiliation{NASA Hubble Fellow}
\affiliation{MIT Kavli Institute for Astrophysics and Space Research, Massachusetts Institute of Technology, 77 Massachusetts Ave, Cambridge, MA 02139, USA}
\affiliation{Department of Physics, McGill University, 3600 rue University, Montr\'eal, QC H3A 2T8, Canada}
\affiliation{Trottier Space Institute, McGill University, 3550 rue University, Montr\'eal, QC H3A 2A7, Canada}

\author[0000-0002-3430-7671]{Alexander W. Pollak}
\affiliation{SETI Institute, 339 Bernardo Ave, Suite 200 Mountain View, CA 94043, USA}

\author[0000-0002-4823-1946]{Vishwangi Shah}
\affiliation{Department of Physics, McGill University, 3600 rue University, Montr\'eal, QC H3A 2T8, Canada}
\affiliation{Trottier Space Institute, McGill University, 3550 rue University, Montr\'eal, QC H3A 2A7, Canada}

\author[0000-0002-6823-2073]{Kaitlyn Shin}
\affiliation{Cahill Center for Astronomy and Astrophysics, MC 249-17 California Institute of Technology, Pasadena CA 91125, USA}

\author[0000-0003-2631-6217]{Seth R. Siegel}
\affiliation{SKA Observatory, 26 Dick Perry Avenue, Kensington, WA 6151, Australia}
\affiliation{Perimeter Institute for Theoretical Physics, 31 Caroline Street N, Waterloo, ON N25 2YL, Canada}
\affiliation{Department of Physics, McGill University, 3600 rue University, Montr\'eal, QC H3A 2T8, Canada}
\affiliation{Trottier Space Institute, McGill University, 3550 rue University, Montr\'eal, QC H3A 2A7, Canada}

\author[0000-0002-1491-3738]{Haochen Wang}
\affiliation{MIT Kavli Institute for Astrophysics and Space Research, Massachusetts Institute of Technology, 77 Massachusetts Ave, Cambridge, MA 02139, USA}
\affiliation{Department of Physics, Massachusetts Institute of Technology, 77 Massachusetts Ave, Cambridge, MA 02139, USA}

\author[0000-0002-7076-8643]{Tarik J Zegmott}
\affiliation{Department of Physics, McGill University, 3600 rue University, Montr\'eal, QC H3A 2T8, Canada}
\affiliation{Trottier Space Institute, McGill University, 3550 rue University, Montr\'eal, QC H3A 2A7, Canada}

\shortauthors{Lazda et al.}
\shorttitle{HCO Commissioning Overview}

\begin{abstract}
This work presents commissioning of the Hat Creek Outrigger (HCO), a dual-polarization 256-element radio interferometer that is part of the Canadian Hydrogen Intensity Mapping Experiment Fast Radio Burst (CHIME/FRB) Outrigger project. Driven by a complex radio-frequency interference (RFI) environment that consistently contaminates  $\sim40\%$ of HCO's usable bandwidth, we implement an improved calibration scheme using Gaussian Process Regression to recover complex gain solutions over RFI contaminated channels.
To validate our method, we test the performance of the array using known transients and continuum sources over relevant timescales used for fast-transient research ($\lesssim$ seconds). We find that our updated calibration scheme results in a $\sim1.69\times~\mathrm{to}~1.87\times$ improvement in the array's point-source sensitivity, while simultaneously maintaining noise properties consistent with thermal statistics. As a result, we find that the array performs consistently within theoretical expectations across $\gtrsim80\%$ of HCO's bandpass. We further observe an improvement in the interferometric performance after applying recently developed spatial filtering techniques for RFI mitigation, which rely on accurate calibration solutions for effective removal of unwanted interference. We conclude that our approach provides a valid framework for improving calibration solutions over RFI contaminated channels for large-$N$ interferometric arrays more broadly. Our work motivates future development of more sophisticated techniques to recover astrophysical information in RFI-contaminated channels, departing from the historical practice of discarding them outright. 

\end{abstract}


  
\section{Introduction}\label{s:introduction}

Radio astronomy has entered an era in which an ever growing fraction of observing bandwidth has become contaminated with radio frequency interference (RFI). As a consequence, the historical preference of simply removing contaminated bands is rapidly becoming infeasible as greater amounts of bandwidth become contaminated. This has spurred recent development in algorithmic approaches aimed not only at \textit{identifying} and \textit{excising} RFI on shorter timescales (e.g., \citealt{deller_software_2010,Rafiei-Ravandi_2023,marshman_characterising_2026,schmid_computer_2026}) but also actively \textit{recovering} astrophysical information within these contaminated channels (e.g., \citealt{van_der_veen_signal_2004,Finlay_2023, zhang2024rfidrunetrestoringdynamicspectra,
andrew_spatial_2026,kuiper_tied-array_2026}). \par

Recently, the sub-domain of transient radio astronomy has seen growth in algorithmic development aimed to recover frequency information. Work by \cite{andrew_spatial_2026} has shown that filtering algorithms based on the Karhunen–Loève transform (KLT) can spatially null dominant RFI sources and maximize sensitivity in the direction of the transient, recovering signal in even heavily contaminated channels where the signal was previously buried in RFI and noise. \cite{kuiper_tied-array_2026} implemented the use of ``flat-fielding" for tied-array beam observations, in which each beam is divided by an averaged reference derived from other beams in the pointing, suppressing red noise and broad-band RFI. Applied to transient astronomical data, the technique resulted in a reduction of false positives while performing low frequency observations of PSR J0250+5854. Both examples showcase how novel algorithmic techniques can lead to successful recovery of a significant quantity of astrophysical information, enabling future transient discoveries despite the inevitable increase in complexity of RFI environments. \par

Wide-field radio interferometers have proven particularly successful for detecting new radio transients. These interferometers adopt designs which maximize sensitivity over large fractions of observable sky to enable rapid survey speeds, typically within the context of $21~\mathrm{cm}$ cosmological surveys (\citealt{DeBoer_2017,2022_chime_overview}). However, such a design has also been found to be ideally suited for detecting bright, radio transient events \citep{collaboration_chime_2018}. Within the field of fast radio bursts (FRBs), the Canadian Hydrogen Intensity Mapping Experiment (CHIME) telescope is one such telescope, boasting an instantaneous field of view of $\sim200~\mathrm{deg}^2$. FRBs are short ($\upmu\mathrm{s}$ to ms), energetic bursts of radio emission whose origins, on a population level, are poorly understood (see \citealt{2021SCPMA..6449501X, Petroff_2022} for a review). Since beginning operations, CHIME/FRB has dominated the global discovery rate of FRBs, accumulating over $4000$ FRBs to-date compared to $\leq500$ bursts accumulated by traditional radio telescope designs \citep{collaboration_chime_2018,collaboration_first_2021,collaboration_second_2026}. The success of CHIME/FRB, combined with the commercial availability of hardware components enabling hundreds-to-thousands element interferometric arrays (i.e., large-$N$ arrays), has spurred development in a number of telescope designs which aim to adopt similar survey strategies to maximize the number of detected transients. Examples of such arrays currently under development include CHORD \citep{vanderlinde_lrp_2019,thechordcollaboration2026overviewcanadianhydrogenobservatory}, BURSTT \citep{lin_burstt_2022}, DSA \citep{hallinan+2019_dsa},  CHARTS \citep{CHARTS_design}, CASM \citep{connor_256-antenna_2026} and SKA-LOW \citep{timmerman2026lowfrequencyvlbiskalow}. \par

All of the aforementioned  projects will share the common challenge of mitigating increasingly oppressive RFI environments driven by modern-day technological advancements (e.g., \citealt{Di_Vruno_2023,Bassa_2024}). This is particularly true for the subset of designs that heavily rely on sky information in order to calibrate individual antenna responses \citep[e.g.,][]{2022_chime_overview}. Such calibration is necessary in order to unlock the full sensitivity enabled by coherently summing voltages across all elements in the array. This challenge is compounded by the fact that the modest to large field of views (FoVs) and near-field zones of $\sim$ hundreds of meters of compact interferometers result in increased sensitivity to the horizon, where RFI sources dominate and contaminate the spectrum of interest. As more of the spectrum becomes contaminated, this method of calibration will become increasingly strained. As such, updated calibration schemes are needed to recover solutions over (ever-growing) RFI contaminated bands. This is particularly necessary to maximize the benefits from techniques aimed at recovering astrophysical information in these contaminated channels which assume that individual interferometric elements are properly calibrated prior to filtering (e.g., \citealt{andrew_spatial_2026}).\par

This work presents the commissioning results of one such large-$N$ interferometer: the Hat Creek Outrigger (HCO) telescope. HCO is the third and final Outrigger telescope part of the CHIME/FRB Outrigger array, a network of wide-field interferometers across North America designed to localize FRBs using very-long-baseline-interferometry (VLBI) (see \citealt{collaboration_chimefrb_2025} for a review). In doing so, the CHIME/FRB Outrigger project seeks to establish the largest sample of FRBs localized to $\sim50~\mathrm{mas}$ to-date, necessary to disentangle various proposed progenitor models as well as unlock FRBs as probes of the diffuse intergalactic baryons and cosmology (e.g., \citealt{Zhou_2014,2024ApJ...973..151K,2025NatAs...9.1226C,2025ApJ...991L..25L,collaboration_frb_2025,2025ApJ...979L..21S,2025ApJ...979L..22E,lanman2026,2026ApJ...998...97C}). \par
Motivated by a complex RFI environment which contaminates up to $\sim45\%$ of HCO's allocated bandwidth between $400-800~\mathrm{MHz}$, we introduce an updated calibration strategy to mitigate and recover astrophysical information in channels previously contaminated by RFI. In particular, we focus on how our updated calibration strategy results in significant improvement in sensitivity across HCO's allocated bandwidth, and performance within the context of VLBI-localizing FRBs. Our work builds upon \cite{2022_chime_overview,lanman_chimefrb_2024}, \citealp{nimmo_gbo_nodate} (hereafter referred to as Paper I, II and III, respectively) which focus on the interferometric performance of the CHIME, KKO and GBO telescopes, respectively, as well as \cite{collaboration_chime_2018}. \par

The structure of this paper is organized as follows: In Section \ref{s:design}, we provide a brief overview of the design of the array. In Section \ref{s:calibration}, we introduce the updated calibration scheme to improve daily complex gain solutions. In Section \ref{s:feed_positions}, we introduce improvements in our feed position determination. In Section \ref{s:performance}, we showcase how updating our calibration strategy enables HCO to achieve significant improvement in autocorrelation sensitivity, recovering astrophysical information in channels previously lost due to RFI. In Section \ref{s:vlbi_performance}, we showcase that the improvement in sensitivity in autocorrelation translates to increased performance in cross-correlation sensitivity in VLBI. Finally, we conclude and summarize our results in Section \ref{s:conclusion}.

\section{Design Overview}\label{s:design}

Given that HCO is the final cylinder in the CHIME/FRB Outrigger VLBI array, specifics regarding the signal chain, design, digital backend and  have already been covered in extensive detail in Papers I, II, III and \cite{collaboration_chimefrb_2025}. In light of this, we provide only a brief summary here and refer the reader to these works for specifics regarding the design of the Outrigger systems. \par

\subsection{Site}
\begin{figure}
    \centering
    \includegraphics[width=1.0  \linewidth]{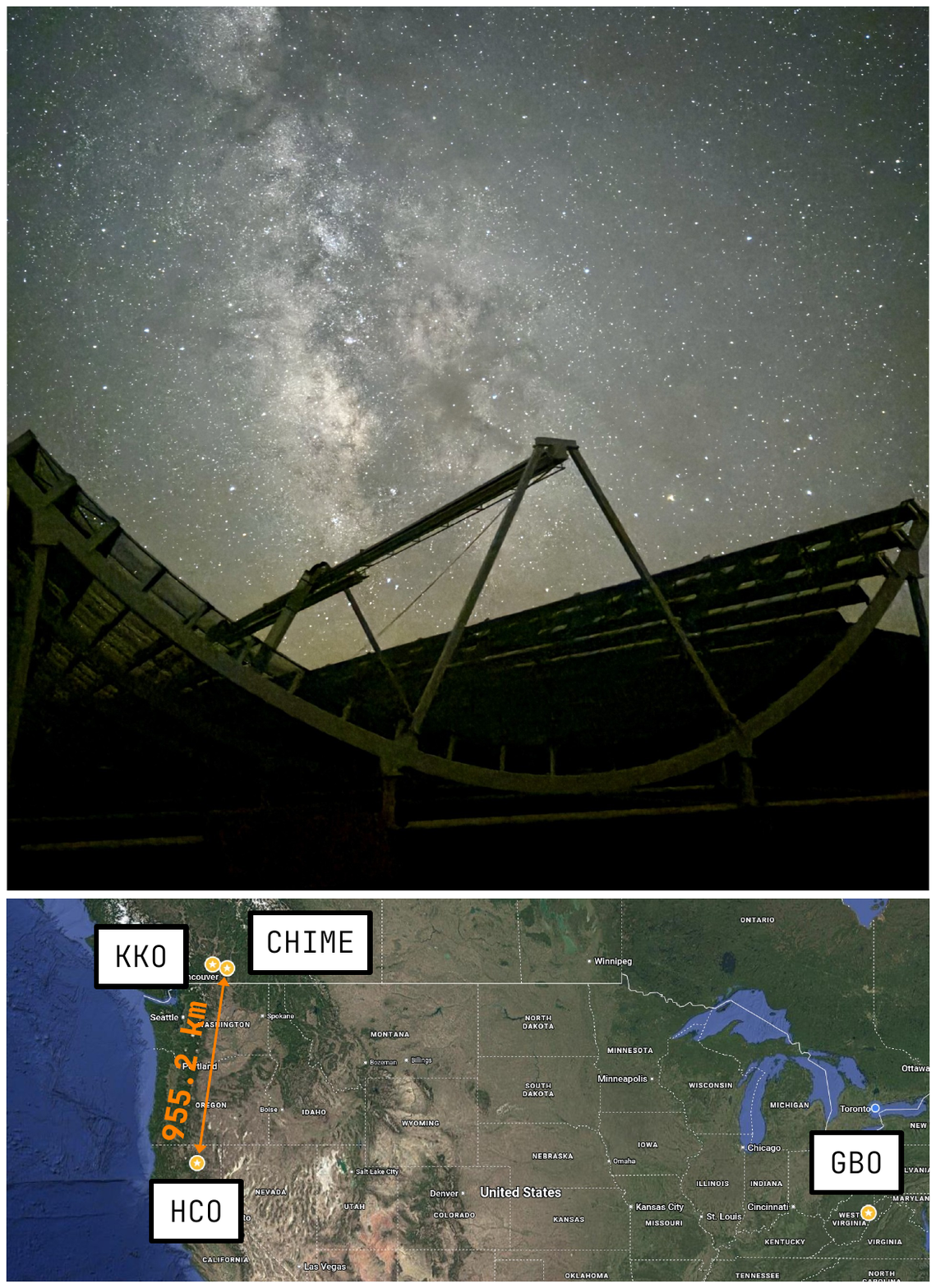}
    \caption{The HCO site. \textit{Top panel:} An image of HCO, located at the SETI Institute in Northern California, a few hundred metres away from the Allen Telescope Array \citep{welch_allen_2009}. \textit{Bottom panel:} Geographic location of HCO relative to CHIME, KKO and GBO. HCO forms a $955.2~\mathrm{km}$ baseline with CHIME, primarily along the North-South axis, enabling 2D localizations of FRBs when combined with KKO and GBO. }
    \label{fig:site}
\end{figure}

HCO is located at the Hat Creek Radio Observatory (HCRO) in northern California, which hosts both the Search for Extraterrestrial Intelligence (SETI) Institute\footnote{\href{www.seti.org}{www.seti.org}} and the Allen Telescope Array (\citealt{welch_allen_2009}). The telescope is located within a valley, offering modest RFI shielding from nearby towns and within 18 km to the nearest town of Burney, California. The telescope as seen at night is pictured in Figure \ref{fig:site}. In Table \ref{tab:site}, we provide specific details regarding the geographic location of the Outrigger. HCO forms a $955.2$-km line-of-sight baseline with CHIME. Located nearly directly south of CHIME, HCO is thus the keystone of the full array, providing astrometric localization precision in the north-south direction (see, e.g., \citealt{collaboration_chimefrb_2025, collaboration_frb_2025}). This contrasts with the KKO and GBO baselines which are nearly collinear and constrain the East-West component of FRB localization ellipses (Papers II, III). \par

\begin{deluxetable}{ll}
\tablecaption{Characteristics of HCO\label{tab:site}}
\tablewidth{0pt}
\tablehead{\colhead{Parameter} & \colhead{Value}}
\startdata
Geographic location & \begin{tabular}[t]{@{}l@{}}40\arcdeg49\arcmin3\arcsec.07N\\
121\arcdeg27\arcmin57\arcsec.66W\end{tabular}\\
Elevation & 1019.34\,m\\
Baseline to CHIME\tablenotemark{a} & 955.2\,km\\
Cylinder width & 20\,m\\
Cylinder length & 80\,m\\
Instrumented cylinder length & 39\,m\\
Cylinder roll & 1.4\arcdeg \\
Cylinder rotation\tablenotemark{b} & 1.22\arcdeg\\
Number of dual-polarization feeds & 128\\
Observing band & 400--800\,MHz\\
\enddata
\tablenotetext{a}{Projected baseline for the target at the CHIME zenith.}
\tablenotetext{b}{East of North.}
\end{deluxetable}

\subsection{Receiver Chain} \label{ss:receiver_chain}

HCO adopts a $20\,\mathrm{m} ~\times~80\,\mathrm{m}$ cylindrical reflector design, marginally rolled by $1.4$ degrees and rotated $1.22~\mathrm{degrees}$ East of North to ensure an overlapping FoV with the CHIME, KKO and GBO telescopes. The cylinder is populated with 128 dual linearly polarized antennas along the inner $39\,\mathrm{m}$ of the focal line relative to the center of the cylinder. A pair of custom LNAs \citep{Bandura_2014,2022_chime_overview} are attached directly to the two polarization outputs of each feed. Each polarization signal is independently amplified and processed. 40-m low-attenuation coaxial cables carry the signals from the focal line to the local compute cluster housed in an RF shielded room within a shipping container. Custom filter amplifiers (FLAs) within the RF shielded room amplify and band-pass filter the signals to 400--800 MHz before they are digitized by the digital backend (Section \ref{ss:baseband_n2}). \par

\subsection{Digital Backend \& Data Products } \label{ss:baseband_n2}
The amplified and filtered analog signals are digitized and channelized by a field-programmable gate array (FPGA)-centered F-engine. The HCO F-engine, as with other CHIME sites, is implemented using the ICE platform \citep{bandura_ice_2016}. The analog signals are sampled by the ICE motherboards' ADCs into 8 bits at 800 MSPS yielding a bandwidth of $400~\mathrm{MHz}$ (Paper I). 
These data are hereafter referred to as ``raw ADC" data, with a native time resolution of $1.25~\,\mathrm{ns}$ per sample. While the full stream of raw ADC frames undergo further processing, a small subsample of raw ADC data (1 frame of 2048 samples for all $256$ inputs) is recorded every $30~\,\mathrm{s}$. These data are primarily used for assessing input health via simple root-mean-square (RMS) diagnostics and for assessing the overall RFI environment at HCO (see Section \ref{ss:rfi_environment}). One F-engine input is connected to the 10 MHz signal of a rubidium clock; the raw ADC frames of this input are saved at a cadence of $0.2\,\mathrm{s}$. These data will be used to perform high precision clock corrections  within the context of VLBI localization of FRBs \citep{Mena_Parra_2022, collaboration_chimefrb_2025}.\par

The digitized data from each input are then passed through a polyphase filter bank (PFB) that separates the bandwidth into 1024 frequency channels with $390.625\,\mathrm{kHz}$ frequency resolution, $2.56\,\upmu\mathrm{s}$ time resolution. They are then re-scaled using computed digital gains and then converted into $4+4j\,\mathrm{bits}$. Channelized data from each input is then sent to a GPU-based $X$-engine. The $X$-engine produces two data products in real time: cross-correlation visibilities between each pair of antennas (``$N^2$ data'') and a 40-second ring buffer that holds complex baseband data for each frequency. We summarize each of the saved data products post digitization stage below. 

\paragraph{$N^2$ data} The first data product outputted by the $X$-engine are autocorrelation and cross-correlated visibilities across all $N = 256$ input pairs. The data have a time resolution of $40\,\mathrm{s}$ and frequency resolution of $390.625\,\mathrm{kHz}$. The data are recorded to disk in real-time in \textsc{hdf5} file formats in $30$ minute chunks. These data are saved continuously throughout the commissioning phase but will be compressed in the future by only saving data during calibrator transits. The coarser time resolution of $N^2$ data results in the data primarily being used in analyses which benefit from longer integration times. These analyses typically revolve around calibration and determining feed positions (see Section \ref{s:calibration}). 

\paragraph{Baseband data} The second data product outputted by the $X$-engine is channelized voltage data with $2.56~\upmu\mathrm{s}$ time resolution and $390.625~\mathrm{kHz}$ resolution, per input.  Hereafter, we refer to these data as ``channelized baseband" data, or simply ``baseband" data\footnote{We note that this convention is chosen to remain consistent with Papers I, II and III.}. These data represent the primary data products used in the VLBI localizations of FRBs and are used to assess the overall performance of HCO as an interferometer (see Sections \ref{s:performance} and \ref{s:vlbi_performance}).

\subsection{Data Flow \& Transport}\label{ss:triggering}
\begin{figure*}
    \centering
    \includegraphics[width=1.0\linewidth]{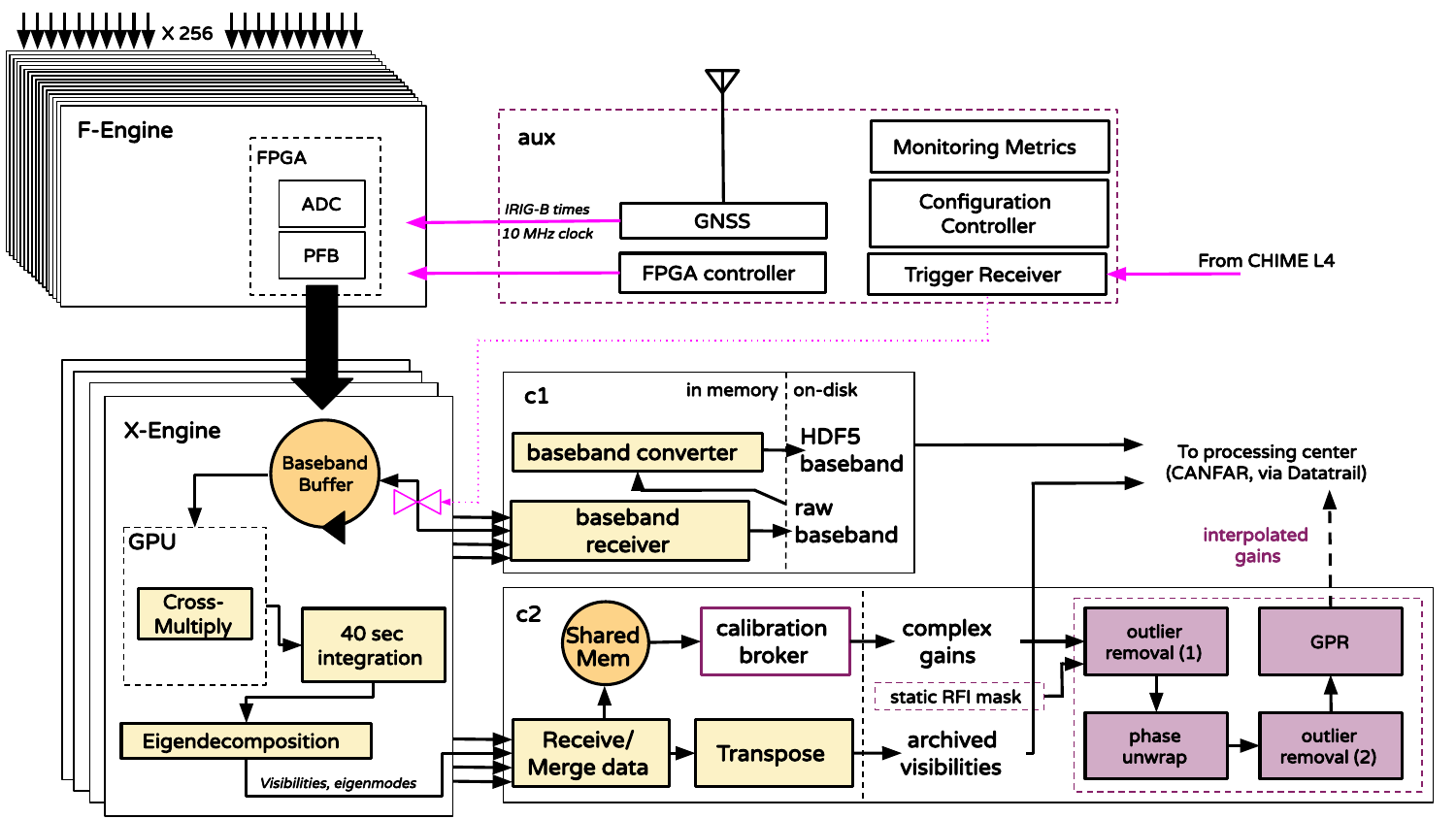}
    \caption{Updated digital system configuration for HCO post analog chain, adapted from the original from Fig. 3 in Paper II. Differences between system configuration for KKO (Paper II) and HCO are indicated by the dashed magenta borders. These changes include (i) consolidating controllers, receivers and monitoring metrics under a single auxiliary node and (ii) an upgraded calibration scheme to recover solutions originally lost from RFI (see Section \ref{s:calibration} for details). Pink lines indicate controllers, while black lines indicate data flow. }
    \label{fig:digital_system}
\end{figure*}

A summary of the digital system configuration post analog chain is visualized in Figure \ref{fig:digital_system}. The overall design draws heavily from previous Outrigger system configurations (see Fig. 3 in Paper II) with two key distinctions. First, services responsible for controlling key hardware components (e.g., the $F$- and $X$- engines), receiving triggers responsible for capturing FRB data and merging monitoring metrics are consolidated to a single auxiliary node, rather than split over two nodes. Second, the Figure reflects how our updated calibration scheme (Section \ref{s:calibration}) is integrated into the broader digital system configuration. \par

Following Figure \ref{fig:digital_system}, voltage data channelized into baseband data are continuously stored in a $40$ second memory ring buffer. Upon detection of an FRB by the CHIME/FRB backend \citep{collaboration_chime_2018}, a trigger is sent to each of the Outrigger sites. Upon receipt of the trigger, the baseband data are written to disk, saving up to a few hundred milliseconds of data surrounding the transient for offline processing. \par

The baseband data are then automatically backed up to \textsc{minoc}, a cloud storage manager, via \textsc{datatrail}\footnote{\href{https://chimefrb.github.io/datatrail/}{https://chimefrb.github.io/datatrail/}}, an in-house built software package designed specifically for tracking, managing and archiving events triggered by CHIME/FRB. The HCO site is equipped with fibre-optic cable internet, allowing for transfer speeds up to 500 $\mathrm{Mb\,s^{-1}}$. Offline processing is carried out on \textsc{canfar} \footnote{\href{https://www.canfar.net/en/
}{https://www.canfar.net/en/
}}, with data from each site pulled locally via \textsc{datatrail}. Further process, including caliration, baseband localization, RFI filtering, and VLBI correlation, are performed at \textsc{canfar} \citep{leung2024vlbisoftwarecorrelatorfast}.  

\subsection{RFI Environment \& Monitoring }\label{ss:rfi_environment}
We wish to assess the overall RFI environment at HCO to identify dominant sources of RFI that contaminate HCO's observing bandwidth from $400-800~\mathrm{MHz}$. Since beginning operations, both HCO and KKO (Paper II) have observed an increase in RFI activity, with both Outriggers recently identifying new unique sources of persistent RFI associated with broadening telecommunication bands. As a consequence, we monitor the RFI environment at HCO on a daily basis to be alerted of any significant changes in the radio spectrum between $400$ to $800~\mathrm{MHz}$. \par
We utilize the saved subsample of raw ADC data (Section \ref{ss:baseband_n2}), channelizing each frame of 2048 samples, per input, into complex voltage measurements with a time resolution of $2.56~\upmu\mathrm{s}$ and $390.625~\mathrm{kHz}$ chunks via a Fast Fourier Transform (FFT). For each sample saved every 30\,s across all inputs, we compute the autocorrelation power as a function of time. Finally, we collapse over the input axis by computing the median power across all powered inputs.\par

\begin{figure}
    \centering
    \includegraphics[width=1.0\linewidth]{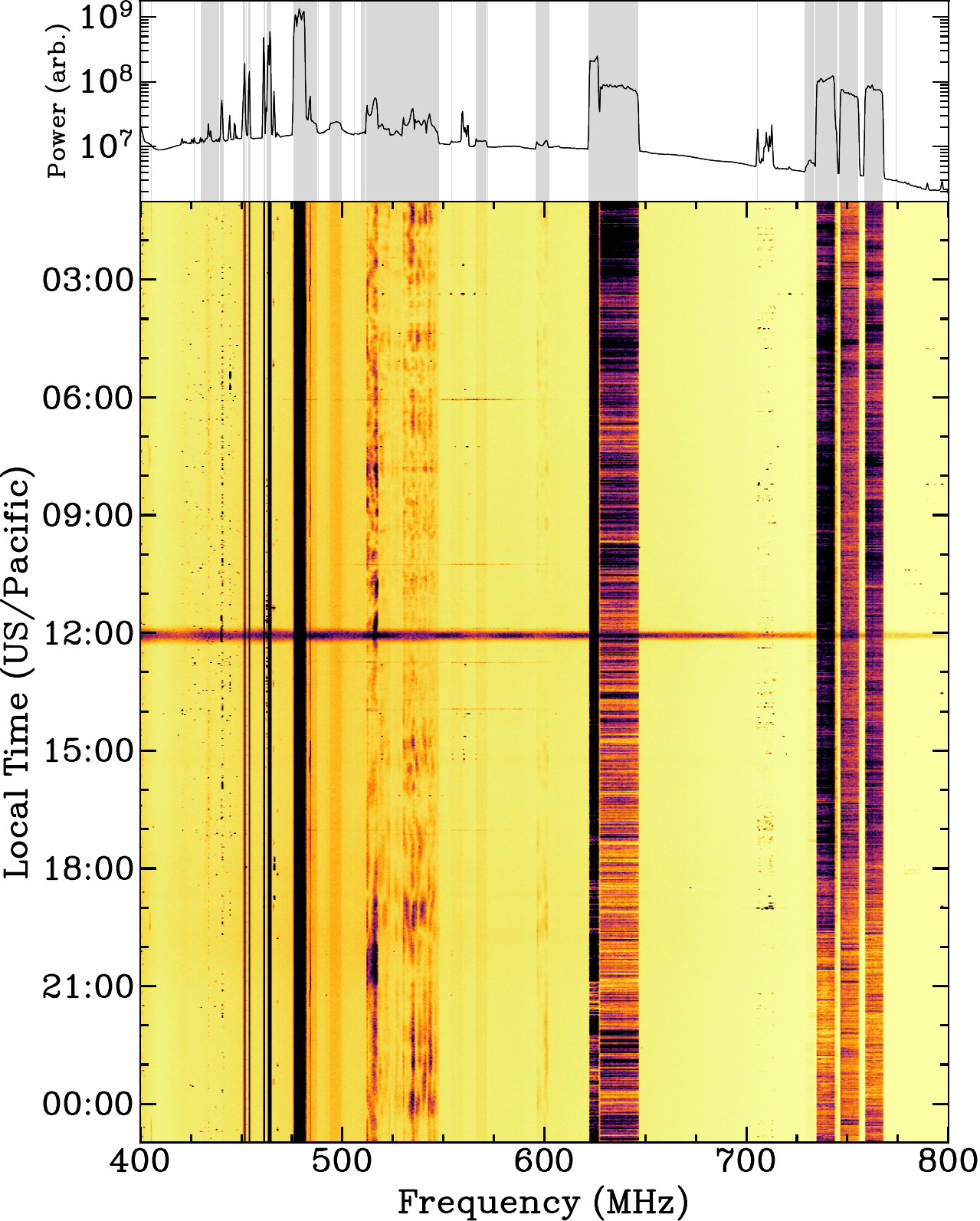}
    \caption{Median 24 hour raw ADC dynamic power spectrum over all powered inputs between MJD 61161 and MJD 61162. Channels exhibiting persistent RFI are indicated by the shaded grey regions in the top panel and derived using the eigenvalues and eigenvectors of the $N^2$ visibility matrices. In total, the mask represents $\sim38\%$ of HCO's allocated bandwidth. Accounting for channels with transient RFI over $24~\mathrm{hr}$ timescales, the fraction of bandwidth contaminated by RFI increases to $\sim45\%$. The Sun transit is clearly visible around 12:00 PT.}
    \label{fig:rawadc_24hr}
\end{figure}

In Figure \ref{fig:rawadc_24hr}, we plot the median autocorrelation power spectrum across all inputs over $24~\mathrm{hr}$ as function of frequency. The Sun transit at 12:00 PT is clearly visible. The power, integrated over time, is indicated in the top panel. We construct a persistent RFI mask by computing the statistics of the visibility eigenvectors over $24~\mathrm{hr}$ timescales following the procedure detailed in section 3.2 in Paper II and plot the result in Figure \ref{fig:rawadc_24hr}. Briefly, we compute the $256$-element vector $\mathbf{a}(t,\nu) = \sum_i \lambda_i \mathbf{e}_i\mathbf{e}_i^*$, where $\mathbf{e}_i$ are the eigenvectors and $\lambda_i$ are the eigenvalues of the visibility matrix (Section \ref{ss:cal_broker}), $i = 0...3$ and known (faulty) inputs are removed from the array prior to computing any statistics. We then compute $N(\nu)$ by computing the median absolute deviation (MAD) over time and summing over inputs. Finally, we flag any channels for which $N(\nu)> \mathrm{mean(\mathbf{a}}) + 4\mathrm{std}(\mathbf{a})$ where the mean and standard deviation are computed over all axes. 

As of June 2026, our $24~\mathrm{hr}$ measurement indicates that $\sim38\%$ of HCO's allocated bandwidth is contaminated by persistent RFI. Repeating this measurement over week- to month-long timescales indicates that this fraction remains broadly constant in time, with day-to-day variation resulting in fluctuations in masked fractions at the $\sim1\%$ level. However, a subset of RFI sources are intermittent (see, e.g., $700-725~\mathrm{MHz}$ in Figure \ref{fig:rawadc_24hr}). Accounting for these transient RFI sources, the fraction of bandwidth contaminated by RFI increases to $\sim45\%$. We discuss the implications of this RFI within the context of calibrating the array in the following section. \par

\section{Complex Gain Calibration}\label{s:calibration}
VLBI cross-correlation is done between CHIME and the Outriggers using \emph{beamformed} baseband data: the voltages are summed across active\footnote{Here, \textit{active} inputs refers to all inputs powered on at the FLA stage. Those powered off represent inputs with known analog issues (e.g., faulty LNAs, loose cable connections, etc.) which result in abnormal voltage measurements. As of August 2026, the number of elements typically powered off is around $\sim30$ out of a total $256$ inputs.} array elements with complex weights that maximize the response to a particular direction.
 This is achievable provided that the arriving signals are aligned (in phase) at each input prior to summing. However, as incoming signals propagate through each of the independent receiver chains introduced in Section \ref{ss:receiver_chain}, systematic delays introduced along the signal chain result in misalignment of the phases between inputs at the digitization stage. Additional misalignment is introduced if the assumed geometric position of each input is incorrect. Depending on the degree of phase-misalignment between each input, not accounting for these contributions can result in significant (and in some cases, complete) loss in sensitivity after summing complex voltages. For this reason, we compute daily complex calibration solutions (also referred to as ``complex point source gains" or simply ``complex gains"\footnote{Note that complex gain calibration (Section \ref{s:calibration}) is not to be confused with digital gain calibration (Section \ref{ss:baseband_n2}). They differ in their ordering of application: the former is applied post digitization stage, whereas the latter is applied as part of the digitization stage. Improving \textit{complex} gain calibration solutions is the focus of this work.}) to remove systematic phase offsets present in recorded voltage data and assess the precision to which each feed's geometric position is known. In Sections \ref{ss:cal_broker}, \ref{ss:GPR} , \ref{ss:gain_comparison} and \ref{ss:gain_stability}, we discuss how these solutions are obtained and improved to mitigate the impacts of RFI. In Section \ref{s:feed_positions}, we describe the method used to determine and correct the geometric positions of each feed in the array.

\subsection{The Calibration Broker}\label{ss:cal_broker}

Obtaining and applying complex gain solutions to downstream voltage data serves two primary purposes: (i) to scale input fluxes from arbitrary correlator units to units of flux density, and (ii) to remove input-dependent instrumental phases. With the stated goal of localizing FRBs, we are primarily interested in obtaining robust and stable phase calibration solutions. Specific details regarding the real-time processing of calibration solutions can be found in Section 2.5 of Paper I. Here, we summarize the key points on how the initial solutions are computed.  \par

Throughout daily operations, integrated $N^2$ visibility matrices (Section \ref{ss:baseband_n2}) are recorded continuously, fringe-stopped towards the known position of a calibrator as it passes through HCO's FoV  (e.g., Cygnus A or Casseiopeia A; abbreviated as Cyg-A and Cas-A hereafter) and spectrally decomposed into their first four eigenmodes using an iterative process \citep{andre_renard_2021_5842660}. In the presence of a bright radio source and in the absence of bright RFI, the matrices become dominated by the first two eigenmodes. These two eigenmodes correspond with each feed polarization.  Since the visibilities have been fringe-stopped, the phase of the computed eigenvectors include only residual systematic contributions. Further, by normalizing the eigenvectors by the known flux density of the source \citep{Perley_2017}, a conversion factor which scales the data from correlator units to Janskys is obtained. Thus, summing the first two (weighted) eigenvectors yields complex gain calibration solutions of the form $\tilde{g_i}(\nu)\equiv|g_i(\nu)|e^{-j\phi_i(\nu)}$, where $|g|$ is the gain amplitude per input $i$ as a function of frequency $\nu$ in units of $[\mathrm{Jy/corr.}]$ and $\phi$ is the input-dependent instrumental phase. The above process is automated through a \textit{calibration broker}, a Python-based service that is provided schedule of calibrator transits and computes calibration solutions on a daily basis (Paper I).\par

We note that given sensitivity limitations, HCO can only reliably obtain calibration solutions using Cyg-A and Cas-A; Taurus A or Virgo A are too faint to robustly calibrate across the entire 400 MHz bandwidth as CHIME does. 

\subsection{Improving Calibration Solutions over RFI Contaminated Channels using Gaussian Process Regression}\label{ss:GPR}

 HCO suffers from an aggressive RFI environment with $\sim{45}\%$ of the bandwidth contaminated to some degree (Section \ref{ss:rfi_environment}, see Fig. \ref{fig:rawadc_24hr}). A significant fraction of this RFI is either contained to narrow frequency ranges, exhibiting brightness comparable to typical sky load, or intermittent. As an immediate consequence, a similar percentage of the raw solutions outputted by the calibration broker (Section \ref{ss:cal_broker}) are also contaminated because neither Cyg-A nor Cas-A is sufficiently bright in these channels to dominate the eigen spectra over the timescales ($\gtrsim 40~\mathrm{sec}$) that the calibration solutions are computed. As a result, the phases of the complex gains in these contaminated channels no longer accurately reflect the true systematic instrumental phase contributions, resulting in these channels becoming effectively unusable if left uncorrected. \par

Fortunately, during the $\sim$milliseconds over which a FRB lasts and is detected by HCO, a large fraction of these frequency channels are not actively saturated by RFI. Consequently, these channels could be recovered if their calibration solutions can be improved. Improving calibration solutions over RFI-contaminated channels would also enable us to use the methods of \cite{andrew_spatial_2026} to suppress RFI. This new technique uses a KLT to spatially null dominant RFI sources in baseband data, maximizing sensitivity toward a transient target. This has been shown to significantly improve sensitivity even in heavily-contaminated channels, but requires valid calibration solutions to work. This motivates our decision to not simply throw away these channels and instead work to actively recover this information. \par

While RFI is also present at the KKO and GBO Outriggers (Papers II, III), the impact of RFI on the calibration solutions is less severe. For example, only $\lesssim20\%-30\%$ of their calibration solutions are contaminated by RFI, while HCO suffers losses up to $\sim45\%$. Thus while KKO and GBO may also benefit from improving calibration solutions over their respective RFI contaminated channels, HCO has significantly more to gain, motivating our analysis. We further highlight that a pipeline designed for improving calibration solutions over RFI contaminated has already been developed within the context of the CHIME cosmology experiment \citep{2022_chime_overview}. However, the adopted scheme uses real-time input and RFI metrics that are finely-tuned for the CHIME instrument to enable 21-cm science to flag RFI prior to interpolation. Since the requirements for transient science are typically less stringent than in a 21-cm cosmological context, we opted for an alternative approach. With the goal of simplifying the calibration process while simultaneously removing our dependence on real-time monitoring metrics, we adopted the strategy of deriving corrections directly from the raw solutions outputted by the calibration broker. In the following sections, we discuss how this task is achieved. 

\subsubsection{Static and dynamic RFI flagging} With the goal of estimating calibration solutions over RFI contaminated channels, we require a framework for estimating which channels are corrupted. We first compute a static RFI mask (see Section \ref{ss:rfi_environment}) to remove persistent RFI. Throughout this analysis, we consider $X$- and $Y$- polarizations independently given that we find that the RFI response is marginally variable between both polarizations. \par

While a static RFI mask removes the majority of RFI over persistently corrupted channels, it is unable to account for intermittent RFI. To account for these channels, we utilize the assumption that $\tilde{g}_i(\nu)$ (as defined in Section \ref{ss:cal_broker}) is expected to evolve smoothly over the bandwidth whereas RFI is expected to rapidly vary. Thus, by computing the derivatives of the gain phases, i.e. $d\mathrm{Arg}(\tilde{g}_i)/d\nu$, we identify sections of the bandwidth which are out of family with respect to the remainder of the bandwidth. Specifically, channels where
\begin{equation}\label{eq:rfi_mask}
    \sigma_{d\phi}(\nu_k) > \mathrm{Median}_{k}(\sigma_{d\phi}(\nu_k)) +\eta\times \mathrm{MAD}_{\nu}(\sigma_{d\phi}(\nu_k))  
\end{equation}
are flagged, where $d\phi_i(\nu_k) = (\mathrm{Arg}[\tilde{g}_i(\nu_k)] - \mathrm{Arg}[\tilde{g}_i(\nu_{k+1})]) $, $k$ indexes neighboring frequency bins, $\sigma_{d\phi}$ is the standard deviation of $d\phi_i(\nu)$ over the input-axis (computed independently per channel), the median and MAD are computed over all frequencies which have not been flagged by the static RFI mask and $\eta$ determines the flagging threshold, which we set to 10 after empirical verification. \par

\subsubsection{Phase un-wrapping and outlier detection}\label{sss:phase_unrwap}
In some instances, the RFI mask defined by Eq.~\ref{eq:rfi_mask} fails to capture outliers where the phase derivatives are smoothly varying but the value extends beyond the general phase trend across the remainder of the band (e.g., jump discontinuities). Since such discontinuities are not expected nor physically motivated, we require that these points are removed prior to interpolation. Broadly, this requires (i) unwrapping the phases, (ii) flattening the phases, and (iii) identifying and removing outliers. Each of these steps is described in detail below. The first two steps are performed independently for each input in the array. \par

A first-order approach to complete the first step is to perform an FFT to identify the dominant source of delay present in the complex gains. However, we found that these solutions were often biased by the remaining RFI channels which we wished to identify, resulting in many instances of the phase failing to be unwrapped. Instead, we fit a linear model to the gain phase of the form
\begin{equation}
   y_{\mathrm{model},i} = 2\pi\nu\tau_i + \phi_{\mathrm{offset},i}, 
\end{equation}
where $\tau$ is the bulk instrumental delay, $\phi_\mathrm{offset}$ is a constant phase offset and $i$  indexes the input axis. We adopt an MCMC approach to sample the posterior distribution to estimate $\tau_i$ and $\phi_{\mathrm{offset},i}$. We assume a Gaussian log-likelihood of the form 
\begin{equation}
    \mathrm{log}\mathcal{L}_i \propto -\frac{1}{2}\sum_\nu\left(\frac{\mathrm{Arg}[\tilde{g}_i(\nu)\cdot e^{-j[y_{\mathrm{model},i}(\nu)]}]}{\sigma_i(\nu)}\right)^2,
\end{equation}
where $\sigma_i(\nu) = \sigma_i$ is the uncertainty on the gain phase which is assumed to be equal across all frequency bins. This decision is intentional to avoid biasing the fit by a small subset of RFI contaminated data points with statistically small uncertainties. Note that this assumption only reasonably holds provided that the majority of the remaining data are RFI-free. This is true in our case since the majority of RFI contaminated channels have been removed via the static RFI mask and Equation \ref{eq:rfi_mask}. As for the priors, given that we expect instrumental residual delays to contribute at most up to a few nanoseconds of delay per signal chain, we adopt a uniform prior on $\tau$ between $\pm5~\mathrm{ns}$ and a uniform prior on $\phi_\mathrm{offset}$ between $-\pi$ and $\pi$. Per input, we perform a 1000 step MCMC with 32 walkers using the \textsc{emcee} package to determine the best-fit phase model, discarding the first 250 samples corresponding to the burn-in period. A typical chain has an acceptance fraction of $30\%$.  \par

To achieve the second step, after coarse unwrapping, we further flatten the phases by subtracting off a third-degree polynomial fit to the phases to remove residual curvature. The result is a set of phase residuals centered around zero. We then flag any frequency channels where these residuals are more than two MADs from the median. \par

We repeat the above process on the amplitudes of the gains, with the only differences being that we (i) skip the MCMC-based unwrapping and (ii) relax the MAD flagging to only exclude extreme outliers extending beyond  $\mathrm{Median} \pm 5 ~\mathrm{MAD}$ after empirical verification. The final RFI mask applied to our gains is thus the union of three masks: (i) the static RFI mask constructed in Section \ref{ss:rfi_environment}, (ii) a dynamic RFI mask based on the behavior of the gain phases and (iii) a dynamic RFI mask based on the behavior of the gain amplitudes. 

\subsubsection{Interpolation via Gaussian Process Regression} 
With a final RFI mask established that is robust against both persistent and intermittent RFI, we apply the mask to the raw gains outputted by the calibration broker. We hereafter refer to this (RFI cleaned) dataset as the ``training" dataset. We interpolate over the real and imaginary components of the gains separately using gaussian process regression (GPR) with the \textsc{scikit-learn} Python package \citep{scikit-learn}. GPR is a statistical technique, often employed within the context of supervised machine learning, which uses prior knowledge to predict outcomes with associated uncertainties \citep{2020arXiv200910862W}. Prior knowledge is encoded in a \textit{kernel} which describes the correlation between two function values evaluated at any two inputs. The choice of kernel therefore depends on \textit{a priori} knowledge about the correlations present in the dataset one wishes to interpolate. The choice to use GPR is motivated in scenarios when a physical process is expected to drive correlations in the observed solutions. In the case of CHIME and the Outriggers, a $30~\mathrm{MHz}$ ripple permeates throughout all calibration solutions (Paper I,II). This arises from the reflections in the cylinder with a focal line height of 5 m which impart a 30 MHz standing wave across the band.  \par

To reflect the known correlations in Outrigger calibration solutions, we adopt a Matérn kernel \citep{porcu2023maternmodeljourneystatistics}, fixing the correlation length scale to $l = 30~\mathrm{MHz}$ and smoothing hyperparameter to $s = 1.5$. Fixing $s=1.5$ assumes that the real and imaginary components of the complex gains are once differentiable \citep{scikit-learn}. We determine this to be a reasonable approximation upon inspection of their numerical derivatives over a clean component of the band. We test the sensitivity of the interpolation scheme to the choice of hyperparameter $l$ and $s$  in greater detail in Section \ref{ss:gain_comparison}. 

To model the underlying fluctuations in the complex gains and correctly scale the Matérn kernel, we multiply the former by a constant kernel whose amplitude is a fitted hyperparameter. An initial guess of the constant value is set to the variance of the real and imaginary components training dataset combined:
\begin{equation}
    \sigma_f^2 = \frac{1}{2}(\sigma_\mathrm{Re}^2 + \sigma_\mathrm{Im}^2),
\end{equation}
where $\sigma_\mathrm{Re}  = 1.4826\times\mathrm{MAD}(y_\mathrm{Re})$ and $\sigma_\mathrm{Im}  = 1.4826\times\mathrm{MAD}(y_\mathrm{Im})$ and $y$ represents the training dataset. Note that the pre-factor of $1.4826$ is included to achieve a robust estimate of the dispersion: the result is similar to computing the standard deviation with the benefit of being insensitive to spurious outliers (Equation 5.44; \citealt{ruppert2011statistics}). We conservatively bound the allowable search range to $\sigma^2 \in[0.01\sigma_f^2,100\sigma_f^2]$. For a typical gain solution, the solution favoring either bound of $0.01\sigma_f^2$ or $100\sigma_f^2$ occurs less than $\leq 1\%$ of the time. With an independent kernel constructed per input, we regress against the (mean subtracted) training dataset using GPR and predict the solutions over flagged channels. The mean of the original training dataset is re-added to the predicted solutions to recover the final interpolated gains.

\begin{figure}
    \centering
    \includegraphics[width=1.0\linewidth]{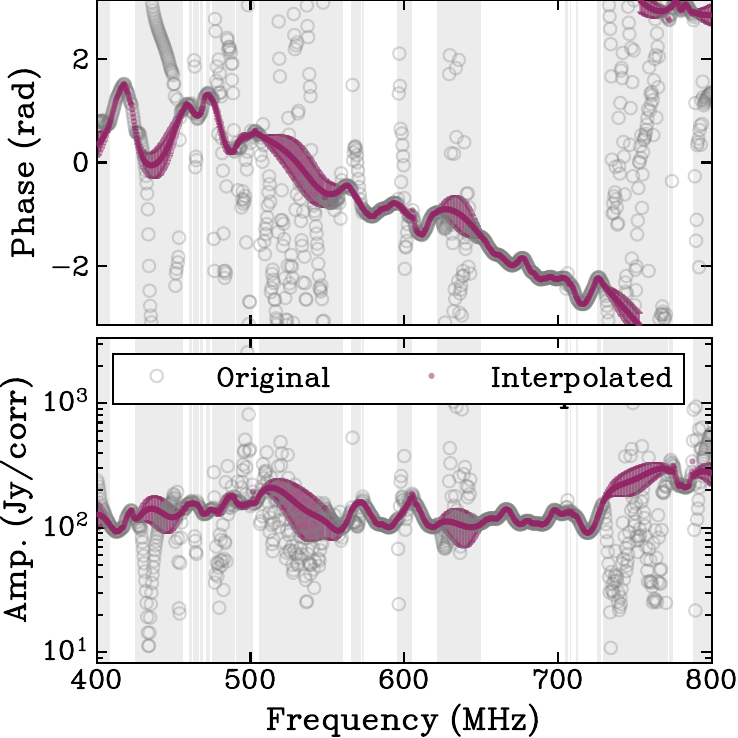}
    \caption{Comparison between an original (transparent grey points) gain solution obtained from a Cas-A transit on MJD 61161 and its reconstructed version (solid purple points) using the methods described in Section \ref{ss:GPR} for a representative input. The final RFI mask prior to interpolation is indicated by the grey shaded regions, representing $60\%$ of HCO's bandwidth. We plot the gain phase and gain amplitude in the top and bottom panel, respectively. Uncertainties on the interpolated values are included, most evident in the $500-550~\mathrm{MHz}$ and $725-775~\mathrm{MHz}$ range. }
    \label{fig:interpolated_gains}
\end{figure}

In Figure \ref{fig:interpolated_gains}, we plot a representative reconstructed gain solution for a single input from a Cas-A transit on MJD 61161. The example emphasizes a number of key points. First, the original solutions are indeed heavily contaminated by RFI: the channels found to be contaminated in Section \ref{ss:rfi_environment} translate to significant degradation in the calibration solutions. We note that our flagging scheme results in some over-flagging ($60\%$ flagged vs. the expected $\lesssim45\%$ as determined in Section \ref{ss:rfi_environment}) due to our choice to aggressively flag phase outliers at twice the MAD of the median. However, the GPR reasonably recovers the solutions over channels lost due to over-flagging, with uncertainties increasing over larger sections of bandwidth. Critically, our scheme ensures that effectively all dominant RFI is flagged and replaced with an interpolated solution. \par 

Second, the interpolator successfully recovers continuous solutions over the flagged channels and avoids introducing rapid variations in phase. However, when no anchor is available (e.g., between $730-770~\mathrm{MHz}$), the interpolation scheme struggles to recover the expected $30~\mathrm{MHz}$ structure seen in uncontaminated parts of the band. Again, however, the uncertainties outputted by the GP interpolator reflect this uncertainty accordingly. \par

Overall, the results in Figure \ref{fig:interpolated_gains} highlight our ability to recover smoothly evolving calibration solutions across the observing bandwidth. Whether the obtained calibration solutions are \textit{valid} is tested in the following section and within the context of transient data in Section \ref{s:performance} and \ref{s:vlbi_performance}. The broad success also further validates our MCMC approach for unwrapping the phases (Section \ref{sss:phase_unrwap}) which we adopt over a naive FFT-based approach. 

\subsection{Sensitivity to hyperparameter selection \& comparison to alternative interpolation schemes}\label{ss:gain_comparison}

In this following section, we determine the impact of choosing to use a GP interpolation scheme with $l = 30~\mathrm{MHz}$ and hyperparameter $s = 1.5$ and compare the results to alternative choices of $l$ and $s$. We also compare the performance of the GP interpolator to two other commonly adopted interpolation schemes: linear and cubic spline interpolators. Our chosen method to compare the performance of each interpolation scheme is discussed below. \par

First, we begin by selecting all \textit{raw} gain solution channels from a single day between $650-700\,\mathrm{MHz}$ as this corresponds to the largest contiguous section of HCO's observing bandwidth free of RFI (Figure \ref{fig:rawadc_24hr} and \ref{fig:interpolated_gains}). Second, we remove channels in one of two ways: the first method randomly removes channels, the second method symmetrically removes channels to both the left and right of $675\,\mathrm{MHz}$ (the mid-point of the clean band). Frequency channels are removed until all $50\,\mathrm{MHz}$ of data have been removed. This is done to emulate RFI which can either be randomly contained to narrow sections of bandwidth, or occupy larger contiguous sections. Each time a certain amount of bandwidth is removed, we interpolate the real and imaginary components as follows. \par

To compare the performance of our chosen GP parameters versus alternative choices of $l$ and $s$, we first fix $l = 30~\mathrm{MHz}$ and vary $s=0.5,1.5$ and $2.5$ to test the sensitivity to our choice of hyperparameter. The range of $s$ corresponds to commonly adopted values that depend on the numerical properties of the interpolant \citep{scikit-learn}. Then, fixing $s = 1.5$, we vary the correlation length between $l = 5\,\mathrm{MHz}$ and $l = 55\,\mathrm{MHz}$ in intervals of $5~\mathrm{MHz}$. \par

To compare the performance of our chosen GP interpolation scheme to commonly adopted interpolation schemes, we repeat the above process but replace GPR with linear and cubic spline interpolation. For any choice of interpolation scheme described above, we interpolate over the removed frequency channels and compute 
\begin{align}\label{eq:gain_err}
    \mathrm{err}_\mathrm{Re} &= \mathrm{Median}_{\nu,i}\left[\frac{|\mathrm{Re}(g^\dagger_i(\nu)) - \mathrm{Re}(g_i(\nu)|) }{|g_i(\nu)|}\right],\\
    \mathrm{err}_\mathrm{Im} &= \mathrm{Median}_{\nu,i}\left[\frac{|\mathrm{Im}(g^\dagger_i(\nu)) - \mathrm{Im}(g_i(\nu))| }{|g_i(\nu)|}\right],
\end{align}
where $g^\dagger$ represents the interpolated gain solutions, $g$ are the raw calibration solutions and the median is computed over all inputs and interpolated channels. \par

\begin{figure}
    \centering
    \includegraphics[width=.85\linewidth]{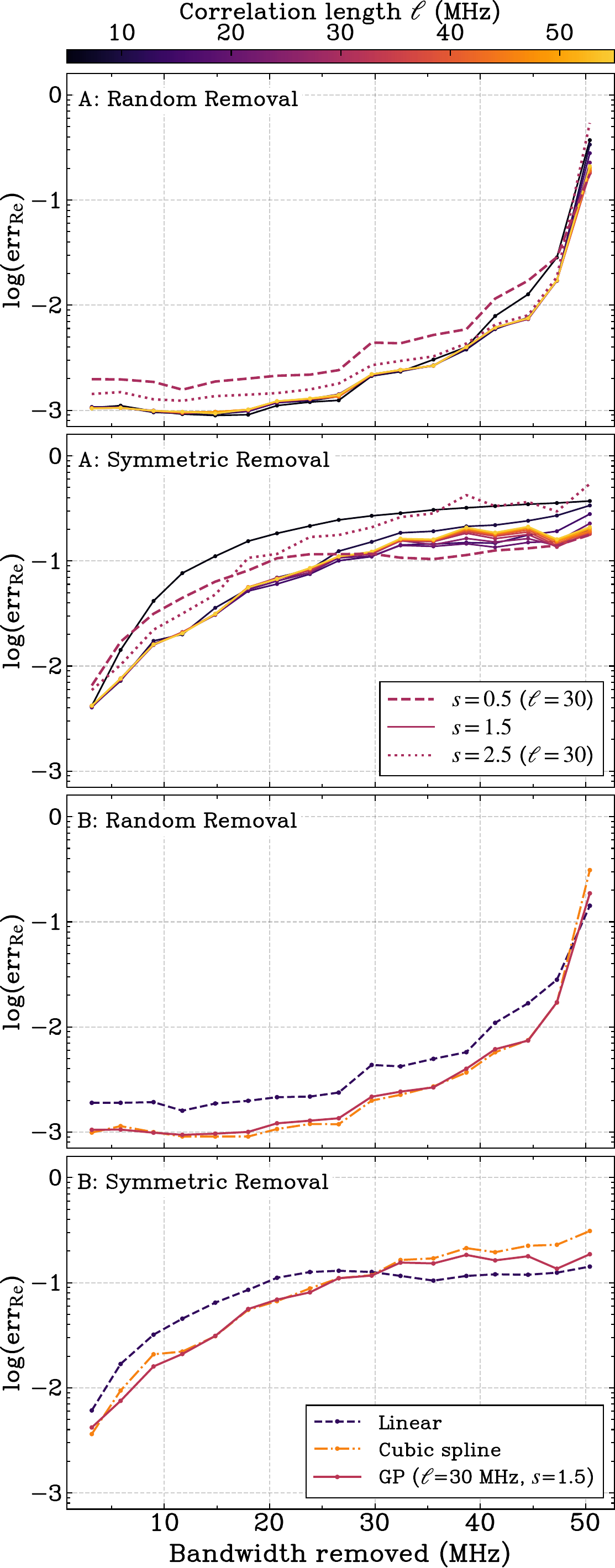}
    \caption{Comparison in performance between various interpolation schemes. In every panel, we compute the error in the real interpolated gain component following Equation \ref{eq:gain_err}. The imaginary component is not plotted for visual clarity, but exhibits near-identical behavior. \textit{(A):} We plot the performance of the GP interpolator for varying hyperparameters ($s\in[0.5,1.5,2.5]$ for fixed correlation length $l = 30~\mathrm{MHz}$). We also compare the performance for varying $l\in[5,55]$ for fixed $s = 1.5$. \textit{(B):} Performance comparison between linear, cubic spline and GP interpolators.  \textit{Random Removal:} Frequency channels are randomly removed between $650-700\,\mathrm{MHz}$ until all $50\,\mathrm{MHz}$ have been removed. \textit{Symmetric Removal:} Frequency channels are symmetrically removed about $675~\mathrm{MHz}$ until all $50~\mathrm{MHz}$ has been removed. } 
    \label{fig:performance_gp_interp}
\end{figure}

We plot the comparison in performance over all of the above configurations in Figure \ref{fig:performance_gp_interp}. While the results highlight only the error in the real component, we note that the imaginary component exhibits near identical  behavior and removed to avoid redundancy. For fixed $l=30~\,\mathrm{MHz}$, $s= 1.5$ performs marginally better across both removal schemes and bandwidth removed in comparison to $s = 0.5$ and $2.5$. Additionally, fixing $s = 1.5$, we find no strong dependence on the choice of $l$ until small correlation lengths are considered, at which point the error can increase to $2\times$ the reference value. We conclude that our choice to fix $l = 30~\mathrm{MHz}$ and $s=1.5$ over alternative choices of $l$ and $s$ is reasonable.  \par

Next, we compare the GP performance (fixed $l= 30\,\mathrm{MHz}, s=1.5$) to both a standard linear interpolator and cubic spline interpolator (see Figure \ref{fig:performance_gp_interp}, bottom two rows). While the linear interpolator performs systematically worse over nearly all trials, the cubic spline and GP achieve similar performance. However, the GP solutions have the additional benefit that they return associated uncertainties on the interpolated values (see, e.g., Figure \ref{fig:interpolated_gains}). While our current version of the VLBI pipeline does not make use of these reported uncertainties, these weights could be useful for (1) optimizing future calibration algorithms which require an initial guess of the calibration solutions, (2) arrays interested in flux calibrating their arrays in the presence of RFI, or (3) optimizing sensitivity at the beamforming stage. Since we aim to explore all three of these scenarios in the future, this motivates our decision to adopt the GP framework.\par

We estimate the uncertainty on the phases of the gains via $\sigma_\phi \approx \left(\frac{1}{2}\mathrm{err_\mathrm{Im}^2 + \frac{1}{2}err_\mathrm{Re}^2}\right)^{1/2}\approx \mathrm{err_\mathrm{Re}}$.  Based on the results in Figure \ref{fig:performance_gp_interp} we expect phase errors on the order of $\sigma_\phi \sim 10^{-3}~\mathrm{rad}-0.3~\mathrm{rad}$ and fractional amplitude variations on the order of $\sigma_{|g|}/|g|\sim0.1\%-30\%$. However, this depends on the degree of bandwidth separation, as well as the assumption that the data surrounding the removed band are free of RFI. In the following section, we test whether the stability of our calibration solutions are dominated by interpolation uncertainties or instrumental contributions. 

\subsection{Stability of Complex Gain Solutions}\label{ss:gain_stability}
Motivated by Papers I and II, we expect gain solutions to remain largely stable over week to month-long timescales. We therefore test the stability of our gain reconstruction scheme (Section \ref{ss:GPR}) to understand the impact of GPR on our ability to routinely recover robust solutions. \par
Following the method outlined in Sections 4.2.1 and 4.2.2 in Paper I, we estimate the variations in our obtained solutions over three weeks of interpolated calibration solutions obtained between MJD 61161 and MJD 61182 using Cas-A. Selecting a reference gain, we compute the relative phase difference and fractional amplitude variation per input as a function of frequency for all gain solutions over the three week time period. We collapse over the time axis by computing the root-mean-square over time per input and frequency. Finally, we collapse over the input axis by calculating the median, 16th and 84th percentile to construct the 68\% confidence interval over inputs.\par

\begin{figure}
    \centering
    \includegraphics[width=1.0\linewidth]{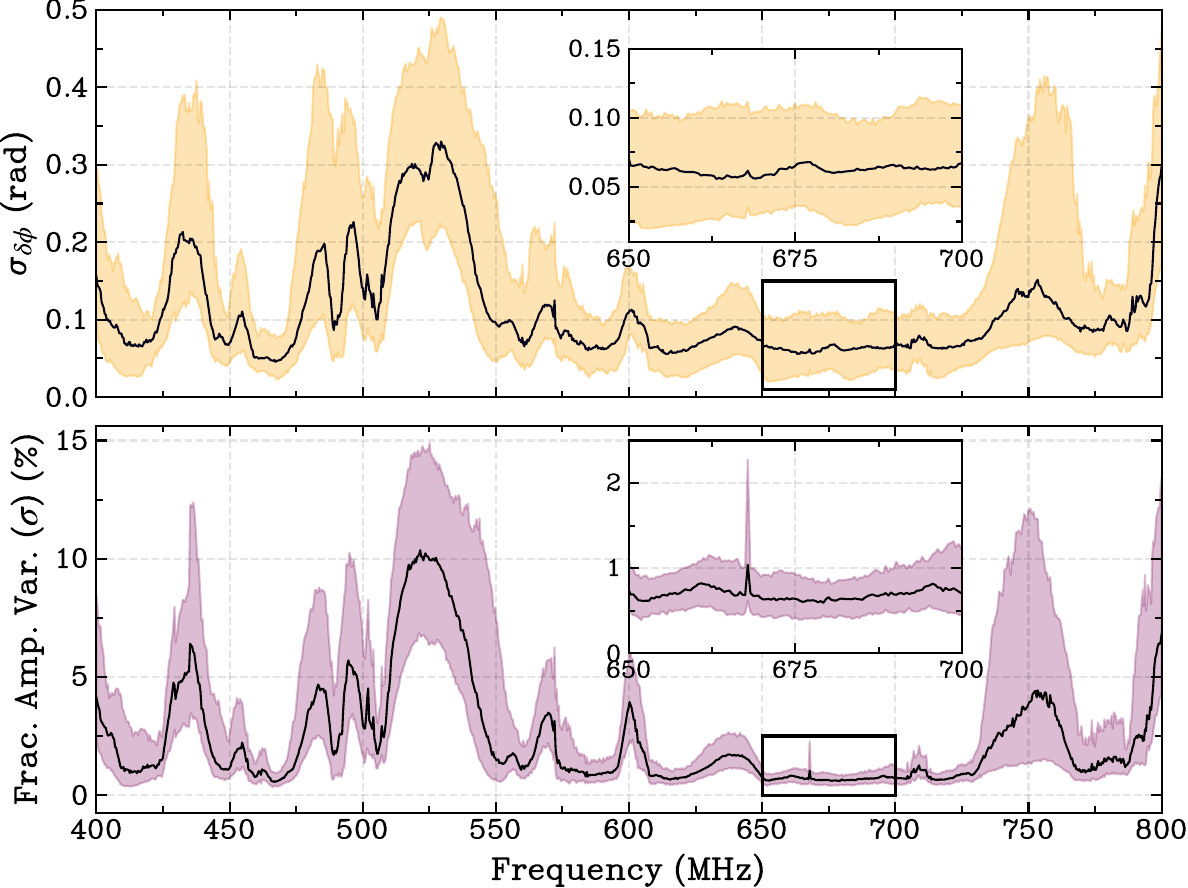}
    \caption{Variation in reconstructed Cas-A gain solutions over the course of three weeks between MJD 61161 and MJD 61182, referenced to a gain solution on MJD 61171.  \textit{Top panel:} We plot the median (solid, black line) standard deviation in phase difference across all inputs over three weeks, excluding inputs with known hardware failures. The orange shaded region represents the region bounded by the $16^{\mathrm{th}}$ and $84^{\mathrm{th}}$ percentile computed over all inputs. \textit{Bottom panel:} Same as the top panel, but instead plotting the variation in fractional amplitude differences. In both panels, we plot zoom-in regions to showcase the behavior of the solutions over RFI-uncontaminated sections of the band. }
    \label{fig:gain_stability}
\end{figure}

A summary of our results is visualized in Figure \ref{fig:gain_stability}. Over subbands that are relatively free of RFI (e.g., $650-700$ MHz), the median variation in phase differences lies slightly above $\sigma_{\delta\phi} = 0.05~\mathrm{rad}$ and fractional amplitude variations falling between $0.5\%-1\%$. These values are broadly consistent with results obtained using CHIME (see Fig. 23 \& 24, Paper I), yielding (raw) phase variations up to $\sigma_{\delta\phi} = 0.03~\mathrm{rad}$ and amplitude variations at the $\sim1\%$ level. Considering instead RFI contaminated bands, we find that both amplitude and phase variations begin to increase by up to an order-of-magnitude, with median phase variations extending to $0.3~\mathrm{rad}$ and median fractional amplitude variations rising to $1-10\%$. Bands which experience the largest variation are associated with sections of bandwidth with aggressive RFI contamination (e.g., $500-550~\mathrm{MHz}$). This behavior is consistent with what was observed in Section \ref{ss:gain_comparison}. We therefore conclude that the primary source of variation over interpolated bands arises from uncertainty in the interpolation. Future improvements could in principle seek to reduce this variation, depending on the scientific requirements by using the outputted solution as an initial estimate of the optimal complex gain solution. Coupled with more sophisticated calibration schemes (e.g., \citealt{2010ISPM...27...30W,adrian_redundantcal,2026arXiv260206109P}), this may allow for more robust and stable solutions to be obtained.  \par

The results of Figure \ref{fig:gain_stability} have two immediate consequences relevant for routinely localizing FRBs. First, under the assumption that our reconstructed gain solutions are correct, the solutions are sufficiently stable across our bandwidth to enable the Karhunen-Loève RFI filter described in \cite{andrew_spatial_2026}. Specifically, the filter is known to be robust to phase calibration and amplitude errors below $\sigma_{\delta\phi} \lesssim 1~\mathrm{rad}$ and $\sigma_{|g|}/|g| \lesssim 1$, respectively, a specification which our reconstructed solutions meet (Figure \ref{fig:gain_stability}). We empirically verify \textit{a posteriori} whether our assumption about gain correctness is valid in Section \ref{s:performance} and Section \ref{s:vlbi_performance}. \par
Second, the stability also has an immediate implication on day to day operations. While the intention is to obtain calibration solutions on a daily basis, there are a number of factors that prohibit HCO from being able to obtain robust calibration solutions every day of the year. These factors range in their severity, where in extreme cases, software failures during calibrator transits can result in missing calibration solutions over week-long timescales. Fortunately, the slow evolving solutions imply that we can use calibration solutions within a few weeks of an event and still successfully calibrate the array. The expected sensitivity loss after beamforming $(\mathrm{S/N})_\mathrm{BF,loss}$ from using a gain solution obtained within three weeks of an event can be estimated by computing \citep{andrew_spatial_2026}: 
\begin{equation}\label{eq:snr_loss}
        \mathrm{(S/N)_\mathrm{BF,loss}} \approx \left(1 - \left[e^{-\sigma^2} + \frac{1-e^{-\sigma^2}}{N}\right]\right)\times100\%,
\end{equation}
where $N$ is the number of active array elements, and the equation is evaluated for both the variance ($\sigma^2$) in phase difference and fractional amplitude variation. Note that this equation assumes that RFI is not present, and that conventional beamforming methods are employed. However, KLT filtering methods can significantly improve the sensitivity when RFI is present \citep{andrew_spatial_2026}.\par
Based on the results of Figure \ref{fig:gain_stability}, this implies an expected sensitivity loss of $\lesssim1\%$ over non-interpolated channels and $1-10\%$ over interpolated channels, respectively, due to phase and amplitude variations. As a result, we maximize the number of FRBs localizable by HCO even in the absence of same-day calibration solutions. 

\section{Feed Position Determination}\label{s:feed_positions}

We motivated the necessity of complex gain calibration because it enables the coherent sum of signals across inputs by removing per-input instrumental delays. However, this is only true if the underlying geometric positions of the feeds are known to sufficient accuracy and precision. If the feed positions are incorrect, coherently summing voltages towards a target sky position will result in the pointing being systematically offset from the intended position. Depending on the degree of error in these positions, this can lead to degraded sensitivity which can vary as a function of pointing. For this reason, we perform a test to validate and correct the geometric positions of each individual input in the array. We note that our method described below deviates from the method introduced in Section 3.5.1 of Paper II. This method, which relied on inferring sky positions based on ratio of eigenmodes between Cyg-A and Cas-A, was unable to probe separation scales $\lesssim10~\mathrm{cm}$, which was believed to be sufficient at the time of commissioning KKO. However, as we will discuss, the original strategy led to sub-optimal performance in the extremities of HCO's beam. As such, we adapted our strategy to probe scales below this limit.\par

\begin{figure*}
    \centering
    \includegraphics[width=1.0\linewidth]{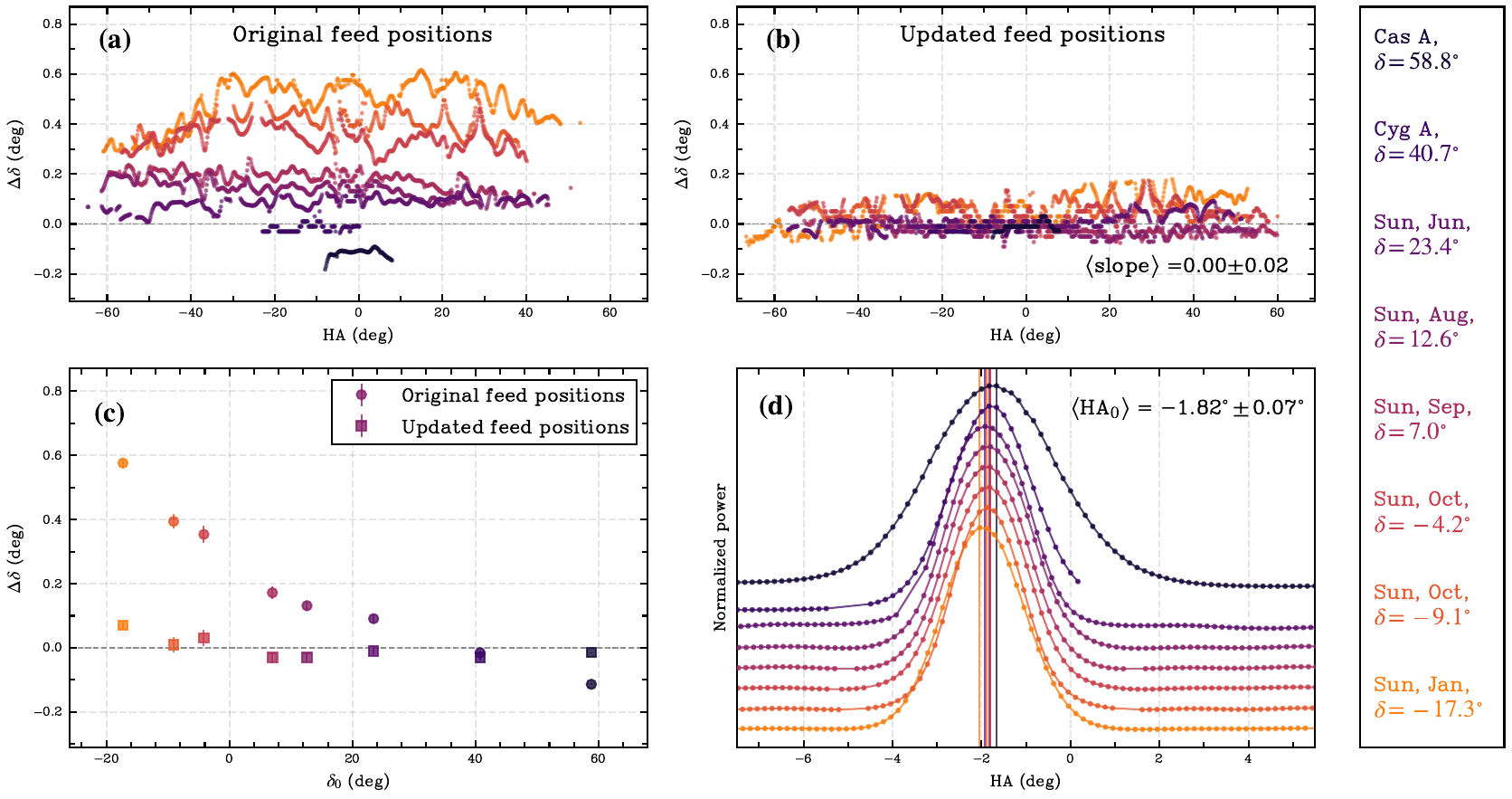}
    \caption{Assessment of telescope orientation and feed position models using observations of bright radio sources (Cas-A, Cyg-A and the Sun). {\it Panels (a) and (b)}: show the measured declination offsets ($\Delta\delta = \delta_\mathrm{meas} - \delta_0$) of Cyg-A, Cas-A and the Sun during their transits across the primary beam and side-lobes. {\it Panel (a)} shows the results for the original feed positions, and {\it Panel (b)} shows the results for the updated feed position model, in which the spacing between successive feeds is reduced by $1.7\,$mm. {\it Panel (c)}: shows the measured $\Delta\delta$ for each source during its transit across the primary beam (average offset over $-4.34^\circ\leq \mathrm{HA}\leq0.66^\circ$). Circular markers correspond to the original feed positions, and square markers to the updated feed position model. The updated model provides accurate source positions over $\sim80^\circ$ declination range. {\it Panel (d)}: shows the visibility power profile (Stokes I) for each source during its primary-beam transit. The brightness peaks of all sources align within $0.07^\circ$, indicating accurate rotation (yaw) of the HCO cylinder. The average peak position ($\langle \mathrm{HA} \rangle =-1.82^\circ$) is also consistent with the longitude difference between HCO and CHIME, indicating accurate roll position of the HCO cylinder.}
    \label{fig:feedpos2}
\end{figure*}

To begin, an initial estimate of the input positions was obtained based on the structural design of the telescope. We estimated the relative $(x,y)$ positions based on the roll and rotation of the cylinder (Table \ref{tab:site}), as well as the separation of the antennas installed within their respective cassettes of $305~\mathrm{mm}$ (Paper I, II). Although HCO feeds and cassette structures are identical to KKO/GBO and we followed the similar installation procedures, small inaccuracies in the positions of the antennas are expected/likely to arise. To verify and correct any systematic offset arising from the installation process, we use calibrated $N^2$ visibilities of astrophysical sources with known astrometric positions. We selected the brightest radio sources for this analysis: Cyg-A, Cas-A and the Sun, and track the position of each source during its transit across the primary beam of the telescope. The extreme brightness of the Sun  allows it to be detected well beyond the primary beam and into the ``sidelobes" (see, e.g., Fig. 9 in Paper II). As such, we continue to track the Sun up to $\pm60^\circ$ into the sidelobes to achieve the statistical power necessary to also constrain the rotation of the cylinder.\par

For each native time bin of $N^2$ data (Section \ref{ss:baseband_n2}) we apply our updated per-input complex gain solutions (Section \ref{ss:GPR}) and fringe-stop the visibilities to $(\alpha_0,\delta)$ for $\delta\in [\delta_0\pm 1^\circ]$, where ($\alpha_0$,$\delta_0$) is the known position of the source\footnote{The tied-array beam of the HCO cylinder station is $\sim0.9~\mathrm{deg}\times7~\mathrm{deg}$ at $600~\mathrm{MHz}$, elongated in the direction perpendicular to the focal line. Since the HCO cylinder is orientated almost parallel to North - South line (rotated $1.22~\mathrm{deg}$ East of North), the telescope's resolution in right ascension is poor. As a consequence, we perform the position checks using declination offsets only.}. We add $X$ and $Y$ polarizations to obtain Stokes-I flux, and average from 650 to 700 MHz, which is the largest contiguous uncontaminated section of the bandpass (see Figure \ref{fig:gain_stability}). The result is a baseline-and-frequency averaged, one-dimensional image. 
Fitting a Gaussian to the image, we measure the observed declination. We repeat this process across all images obtained during the transit and discard any measurements for which the fit failed to converge. Using transits of the Sun throughout the first year of commissioning, we increase our declination coverage to $-17^\circ\leq\delta_0\leq23^\circ$. \par

In Figure \ref{fig:feedpos2} (a) we plot the offset between the measured and true effective source declination: e.g., $\Delta\delta = \delta_\mathrm{meas} - \delta_0$. The results of these measurements for each source are shown in different color track and are obtained assuming our original feed positions that are based solely on the site survey and design. The results clearly show that a systematic offset is present, scaling approximately linearly with the distance from the position of the gain calibrator (see circle points in the panel (c) of Figure~~\ref{fig:feedpos2}). 

 The linear relationship between offset and relative sky position is strongly suggestive of a constant offset in feed spacing which may physically be caused by uniform thermal expansion of the focal line or may have been introduced during the installation process. To test this, we introduced a constant horizontal feed offset to each input (fixed across all inputs) and empirically determined the value required to remove the observed linear relationship (see square points in panel (c) of Figure~\ref{fig:feedpos2}). Our best fit solution implies that feeds are $\sim0.55\%$ (or $\sim1.7 ~\mathrm{mm}$ per feed) closer together than predicted, accumulating to $22~\mathrm{cm}$ error over entire focal line. The repeated declination offset measurements post correction are plotted in the panel (b) of Figure~\ref{fig:feedpos2}. Using the updated feed positions returned accurate source position measurements over large range of declinations (and gain calibrator -- target separations). These measurements also highlight that source position stays broadly consistent over large hour-angles (we measure average slope of $\Delta\delta$ over HA to be $0.00\pm0.02$, i.e. consistent with 0), indicating that the model of telescope rotation is sufficiently accurate. The residual structure in the source position tracks, particularly prominent at low declinations, is driven by the beam response which is highly non-uniform. We note that we do not attempt to account for any effects of varying beam response and do not believe that it has any dramatic impact on our overall fit.\par

Finally, we perform beam-mapping of the HCO telescope using the same dataset to confirm the accurate physical orientation of the telescope. In panel (d) of Figure~\ref{fig:feedpos2} we plot the measured Stokes I of visibilities of the sources as a function of hour angle as they transit through the telescope's primary beam. We find that peak of brightness of each source aligns within $0.07^{\circ}$ and the HA of that peak is consistent with the longitude difference of HCO and CHIME within $0.03^\circ$, confirming correct alignment of HCO and CHIME beams and consequently confirming that the rotation and roll of the telescope is sufficiently accurate.

 In summary, using imaging of bright radio sources with $N^2$ visibilities, we confirmed that the orientation of the telescope is consistent with the designed values. Conversely, we found that the physical distance between the feeds is shorter than expected ($\sim1.7 ~\mathrm{mm}$ per feed). This offset is largely believed to be the result of the feed installation procedure, which was designed to ensure that the gap between cassettes (which house the feeds) could not be larger than planned, but did not restrict how close they could be installed. Overall, however, we find that the updated feed position model improves accuracy of fringe-stopping and beamforming of the Outrigger station data, especially for the sources with large separation from the gain calibrator. We highlight the impacts of correcting our feed positions in greater detail in Section \ref{ss:xcorr_pulsars}. Finally, we note that since corrections detailed throughout this section are purely digital, physical repositioning of the feeds along the focal line is unnecessary.

\section{Impact of Updated Calibration Scheme on Interferometric Performance}\label{s:performance}
In the following section, we investigate the impact of our updated calibration and feed position determination scheme described in Section \ref{s:calibration} and \ref{s:feed_positions} on HCO's overall interferometric performance. Specifically, we describe a series of tests performed on the array which allow us to confirm that our changes do indeed result in significant improvement in HCO's overall performance.

\subsection{System Equivalent Flux Density }\label{ss:sefd}
The first test we perform seeks to assess whether the combined sensitivity of individual antennas and receiver chains are within theoretical expectation, given the measured properties of the receiver chain (see Section \ref{ss:receiver_chain}). Following Section 3.4.1 of Paper II, we compute the system equivalent flux density (SEFD) for each input individually and for the beamformed response. The SEFD is the flux a point source at beam center would need to double system temperature \citep{thompson_interferometry_2017}, and provides a useful metric to compare system noise to target brightness, and to identify malfunctioning inputs. \par

For each individual receiver chain, we average 30 minutes of autocorrelation $N^2$ data (Section \ref{ss:baseband_n2}) captured during the night-time while no dominant radio source is in the field. In doing so, we estimate the response of each individual input when the receiver chain is dominated by noise. By subsequently applying per-input gain calibration solutions from a Cyg-A transit obtained from the same day (Section \ref{s:calibration}), we translate the amplitudes from correlator units to Janskys. This results in an estimate of the per-feed $\mathrm{SEFD}_\mathrm{inp}$.\par

We note that while our interpolation scheme results in continuous calibration solutions independently per input, it does not identify any outlier inputs with respect to the remainder of the array. Thus, to remove inputs that are out of family, we construct a simple mask which isolates and removes any inputs whose log-amplitudes deviate from the median amplitude across all inputs by $\pm5~\mathrm{MAD}$. Finally, to assess whether our improved calibration scheme discussed in Section \ref{s:calibration} leads to an improvement in sensitivity across RFI-contaminated channels, we compare the results of the test with the original and improved calibration solutions.\par

\begin{figure*}
    \centering
    \includegraphics[width=0.7\linewidth]{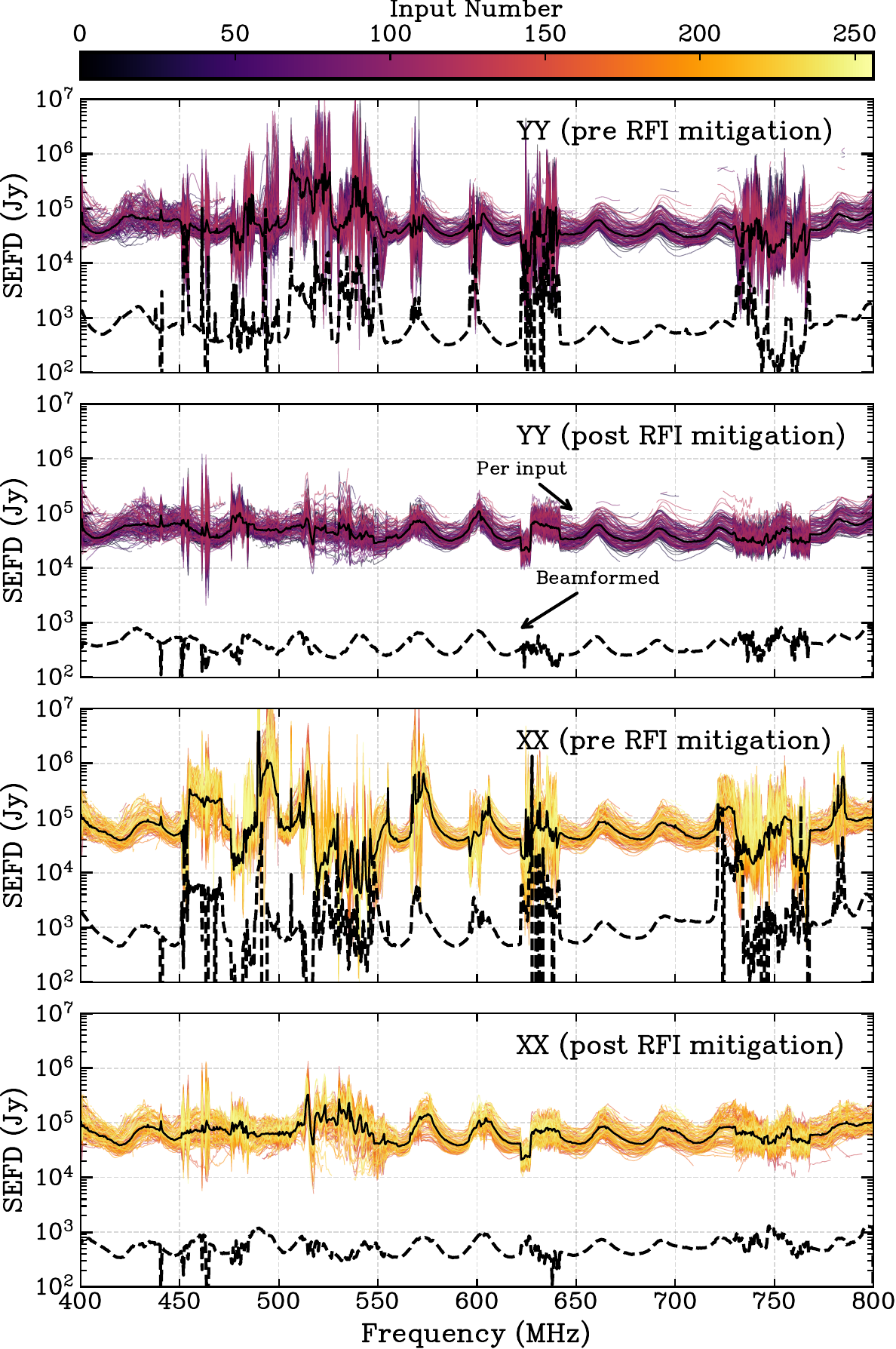}
    \caption{Comparison between per-input and beamformed SEFD measurements for varying calibration configurations. Per-input SEFD measurements (colored) are computed by averaging $30~\mathrm{min}$ of autocorrelation $N^2$ during a night-time transit. Full-array baseband SEFD measurements are computed following Equation \ref{eq:sefd}. We compare the measured SEFD before (pre RFI mitigation) and after (post RFI mitigation) applying our updated calibration solutions, highlighting the improvement in SEFD per polarization after updating our calibration solutions. The median baseband SEFD, reflecting the current instantaneous point-source sensitivity for a source located near the centre of HCO's beam, improves from $990~\mathrm{Jy}$ to $530~\mathrm{Jy}$ and $660~\mathrm{Jy}$ to  $390~\mathrm{Jy}$ in the XX and YY polarization, respectively, after applying our updated calibration strategy (Section \ref{s:calibration}). }
    \label{fig:sefd_test}
\end{figure*}

Individual $\mathrm{SEFD}_\mathrm{inp}$ measurements per input are plotted in Figure \ref{fig:sefd_test}, color-coded based on their polarization. First, for a given polarization, we observe that the individual SEFD measurements \textit{after} applying the updated calibration solutions results in qualitatively smoother variations across the band. This is expected, given that our improved calibration scheme suppresses and smooths solutions over portions of the band that exhibit larger variation driven by RFI. Second, while the original calibration solutions result in median SEFD measurements in excess of $\gtrsim10^{5}~\mathrm{Jy}$ over a significant portion of the band, applying the improved calibration solutions reduces the median SEFD over these same regions to $\lesssim10^5~\mathrm{Jy}$, in-line with previous measurements performed by KKO (Paper II). This reinforces that our improved calibration scheme enables us to achieve close to the expected sensitivity. Finally, comparing the two polarizations, we find that the median SEFD of North-South oriented polarization hands ($YY$) tend to, on average, perform better than the East-West oriented polarization hands ($XX$). This likely suggests that there exists some direction dependence to the observed RFI, or slight variations in response between polarization, resulting in one hand being more susceptible to sensitivity losses despite similar analog designs. \par

While the former test estimates the sensitivity of individual receiver chains, we further wish to estimate the sensitivity of the array as a whole. That is, how sensitive is HCO when voltage measurements across all receiver chains are averaged together? Does our improved calibration technique actually improve the full-array sensitivity? To answer these questions, we compute the \textit{beamformed} SEFD per polarization using a 1.4 second baseband capture. To each of the inputs, we apply our derived complex gain solutions to align each input in phase prior to averaging, again performing the test with and without the improved calibration solutions. In both instances, we assume the \textit{corrected} feed positions to isolate the improvement introduced by updating our calibration strategy. Beamforming to two distinct locations in the beam, we compute the beamformed SEFD following Equation 2 in Paper II: 
\begin{equation}\label{eq:sefd}
    \mathrm{SEFD}_\mathrm{bf} = P_\mathrm{off}(\nu) \frac{F(\nu)}{P_\mathrm{on}(\nu) - P_\mathrm{off}(\nu)},
\end{equation}
where $P_\mathrm{on}(\nu)$ and $P_\mathrm{off}(\nu)$ are the input-and-time-averaged power per frequency channel after beamforming towards Cyg-A and a dark patch of sky, respectively, and $F(\nu)$ is the known flux density of Cyg-A \citep{Perley_2017}. The results are plotted as dashed black lines in Figure \ref{fig:sefd_test}. \par

It becomes immediately clear that the improved calibration solutions result in improvements in sensitivity across both polarizations. In particular, we achieve more smoothly evolving beamformed SEFD measurements across the band, particularly noticeable in the $400-600~\mathrm{MHz}$ band where significant RFI persists at HCO (Section \ref{ss:rfi_environment}). Additionally, we measure a decrease in median SEFD computed over non-flagged frequency channels from $990~\mathrm{Jy}$ to $530~\mathrm{Jy}$ in the XX polarization and a decrease from $660~\mathrm{Jy}$ to $390~\mathrm{Jy}$ in the YY polarization. The improved beamformed SEFD measurements are approximately a factor of $2\times$ smaller than those of KKO, consistent with theoretical expectations based on the assumption that the beamformed SEFD scales with collecting area (Papers II). The values also fall within $5\%~(10\%)$ of theoretical expectations for the $Y$ ($X$) polarization based on the the radiometer equation which states that $\mathrm{SEFD}_\mathrm{bf} = \mathrm{SEFD}_\mathrm{inp}/\sqrt {(N(N-1))}$ where $N$ is the number of inputs used during the beamforming stage.\par 

An alternative framing is to estimate the effective number of inputs recovered by updating our calibration scheme. Defining
\begin{equation}
    N_\mathrm{eff} = \frac{1}{2}\left[\sqrt{1 + 4\left(\frac{\mathrm{SEFD_{inp}}}{\mathrm{SEFD_{bf}}}\right)^2} + 1\right],
\end{equation}
to be the effective number of inputs contributing to the array, we fix $\mathrm{SEFD_{inp}}$ and $\mathrm{SEFD_{BF}}$ to the values before and after updating our calibration scheme.  We measure an increase in $N_\mathrm{eff}$ of $46~(51)$ inputs across the $Y~(X)$ polarization over the baseline performance. We conclude that updating our calibration strategy effectively recovers the equivalent performance of $\sim97$ inputs ($\sim37\%$) across the entire array.\par

Assuming that a detected source is unpolarized and is spectrally uniform across the band, these results imply that our improved calibration solutions yield a median instantaneous sensitivity increase of $1.87\times$ ($\sim87\%$) and $1.69\times$ ($\sim69\%$) in the $XX$ and $YY$ polarization, respectively, in comparison to our original calibration strategy. We note that this improvement is primarily due to a the significant increase in usable bandwidth and recovery of previously contaminated channels which biased the median SEFD to larger values. However, some improvements also arise from the signal in each channel getting systematically brighter. We quantitatively explore this improvement within the context of detected a detected transient in Section \ref{ss:baseband_sensitivity}.  \par

It is not entirely unexpected that by improving our complex gain calibration solutions, we achieve improved sensitivity across the majority of the band. Our results in Figure \ref{fig:sefd_test} confirm our hypothesis introduced in Section \ref{ss:GPR} that the majority of the RFI contamination is sufficiently low power to not completely saturate the baseband data over $\lesssim1.4~\mathrm{s}$ timescales. Instead, the previous loss in sensitivity was simply driven by corrupted calibration solutions. Thus, by applying a second-order correction to align these inputs in phase across the band, we recover significant sensitivity over what were previously lost channels. 

These results further imply that we should expect similar sensitivity improvements with detected transients given that the timescales over which they occur ($\sim 10~\mathrm{ms}$) are significantly shorter than the timescales over which the RFI begins to dominate the signal ($\sim\mathrm{minutes}$ to hours). We test and confirm that this assumption holds in Section \ref{ss:baseband_sensitivity}. \par

\subsection{Radiometer Test }\label{ss:radiometer_test}
Next, we wish to assess the noise properties of baseband data to confirm that the improvements in our calibration scheme result in voltage data that is governed by thermal statistics. First, we assume that the measured voltages are modeled by $V = Ae^{-j\phi} + \epsilon$ where $A$ and $\phi$ encode the amplitude and phase information 
of the measured signal, respectively, and $\epsilon$ represents the noise added to the signal introduced by the receiving system. Second, we assume that $\epsilon$ is drawn from a circularly symmetric complex normal distribution. Our goal is to test whether these assumptions hold across the $400~\mathrm{MHz}$ of bandwidth.\par
Under the aforementioned assumptions, a given time-and-frequency averaged power measurement ($\langle P\rangle  \propto |V|^2$) will have an uncertainty dictated by the radiometer equation \citep{thompson_interferometry_2017}: $\sqrt{\mathrm{Var}[\langle P\rangle ]} = \langle P\rangle /\sqrt{\tau\Delta\nu}$, where $\Delta\nu$ is the bandwidth and $\tau$ is the integration time. The corresponding $\mathrm{S/N}$ of the measured signal therefore scales directly proportional to $\propto\sqrt{\Delta\nu\tau}$. Testing whether the radiometer equation holds over (i) our observing bandwidth and (ii) the timescales over which detected transients occur, is critical to ensure that we can achieve the expected improvements in sensitivity obtained by averaging power measurements both in time and frequency.  \par

We perform the radiometer test using a $500\,\mathrm{ms}$ baseband capture obtained on MJD~61175, the chosen length consistent with the duration of baseband captures currently used to localize FRBs \citep{andrew_astrometry_nodate}. Having calibrated each input with our complex gain solutions, we beamform the array to a dark patch of sky to approximate the response of the array dominated solely by instrument noise. Additionally, we apply the Karhunen–Loève based filter described in \cite{andrew_spatial_2026} to test its impact on the noise properties of the baseband data. We perform the test over $\Delta\nu = 390.625~\mathrm{kHz}$ frequency bins (corresponding to the native frequency resolution of the baseband data products outputted by the $X$-engine; see Section \ref{ss:baseband_n2}) and a timescale of $\tau = 10~\mathrm{ms}$ (corresponding to a typical duration of a captured transient). \par

\begin{figure*}
    \centering
    \includegraphics[width=1.0\linewidth]{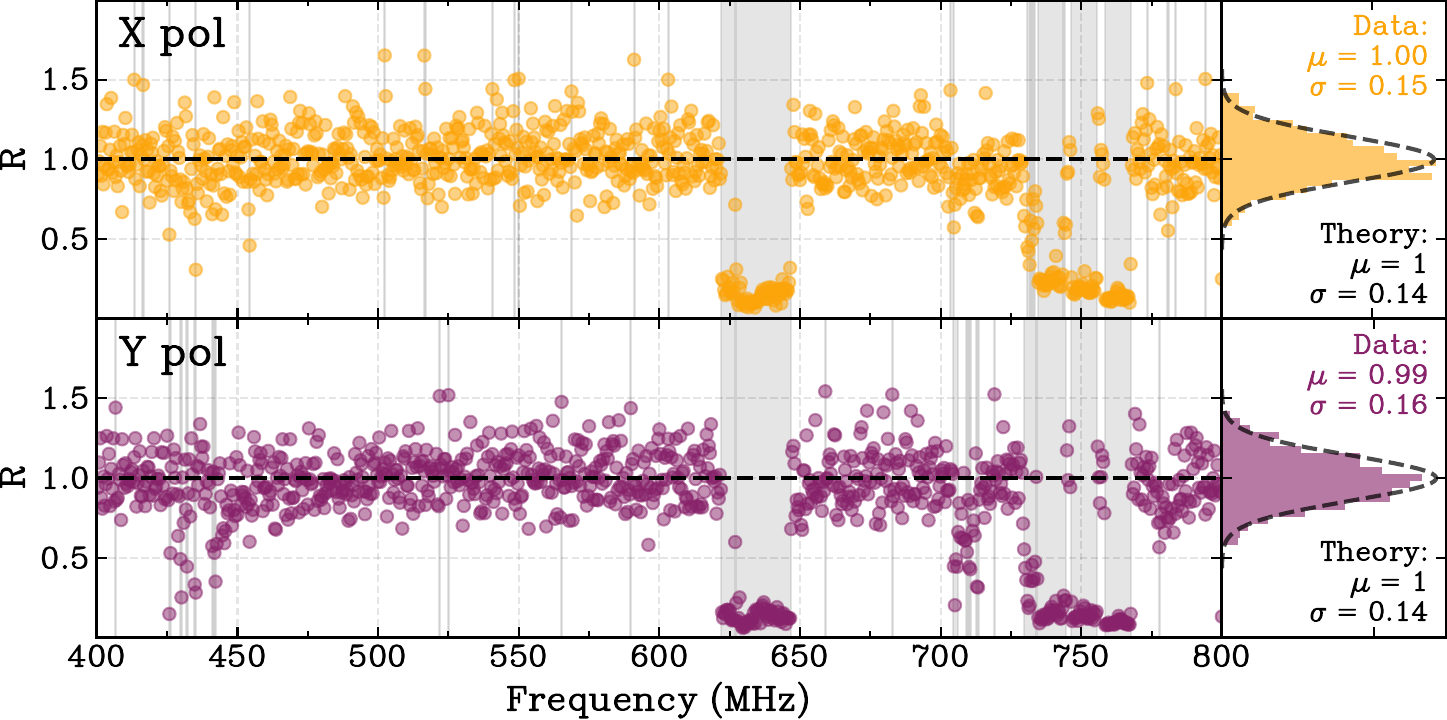}
    \caption{Results of the full-array radiometer test performed, per polarization, using a $500~\mathrm{ms}$ baseband capture using HCO. The baseband data are calibrated using our updated calibration solutions, filtered using the Karhunen–Loève based transform \citep{andrew_spatial_2026} and beamformed to a quiet patch of the sky. We compute the power and average over $\tau=10~\mathrm{ms}$ chunks (broadly consistent with typical duration of captured FRBs) per frequency bin of width $\Delta\nu=390.625~\mathrm{kHz}$. We scale and convert the values to a dimensionless quantity $R$ following Equation \ref{eq:radiometer}. The expectation, assuming that the overall instrumental noise obeys thermal statistics, is highlighted by the dashed black line at $R = 1$.  A histogram of the samples is provided per polarization hand, highlighting that both distributions are broadly consistent with theoretical expectations. Regions where the test fails (see text for failure criteria) are marked by the shaded gray regions and are not included in the histogram. The test succeeds over $83\%$ ($85\%$) of the band for the $Y$ ($X$) polarization. }
    \label{fig:radiometer}
\end{figure*}

In Figure \ref{fig:radiometer}, we plot the results of the radiometer test performed with HCO. Similar to Paper II, we re-cast the radiometer equation to an expected ratio based on the number of samples used to obtain our power measurements: 
\begin{equation}\label{eq:radiometer}
    R \equiv \frac{\langle P\rangle}{\sqrt{\mathrm{Var}[\langle P\rangle ]}}\frac{1}{\sqrt{\Delta\nu\tau}}.
\end{equation}
 In practice, the $500\,\mathrm{ms}$ baseband capture is partitioned into $50$ bins of $10\,\mathrm{ms}$ duration, complex conjugate multiplied to form power, and averaged over each $10\,\mathrm{ms}$ chunk. The mean $\langle P\rangle$ and standard deviation $\sqrt{\mathrm{Var}[\langle P\rangle]}$ are then computed over all $n = 50$ time-averaged power samples.  Assuming Gaussian statistics, we expect $R = 1$ with fluctuations in $R$ inversely proportional to the square root of number of samples used to average over the power (i.e., $\sigma = 1/\sqrt{n} \approx 0.14$ for $n = 50$). We flag any data whose value extends beyond $R = \upmu\pm3\sigma$, where $\upmu = 1$ and $\sigma=0.14$ to identify any channels which do not obey the radiometer equation. \par

We find that over the majority of the band, the statistics of the power measurements are broadly consistent with thermal statistics, with mean values and standard deviations over non-flagged data consistent with theoretical expectation. We find that $\sim17\%$ ($\sim15\%$) of data are flagged over the $Y$ ($X$) polarization. The flagging is primarily concentrated to the $625-650~\mathrm{MHz}$ band and $730-770~\mathrm{MHz}$ band, two known dominant sources of RFI associated with broadband telecommunication services. \par

These two particular RFI bands are of specific interest within the context of the KLT filter \citep{andrew_spatial_2026}. We observe some improvement over these bands after applying the KLT filter, which leads to the recovery of a few channels in the $730-770~\mathrm{MHz}$ band (showcased within the context of a detected transient in Section \ref{ss:impact_interp}). The KLT filter also improves the statistics of a number of channels contaminated with narrow-band RFI present in Figure \ref{fig:rawadc_24hr}. However, the results in Figure \ref{fig:radiometer} largely suggest that the noise properties of the telecommunication bands remain heavily governed by non-thermal statistics. This could imply that our gain solutions over these channels are incorrect, resulting in the filter being unable to adequately spatially suppress the RFI. Alternatively, it could also imply that the properties of the RFI does not obey the assumptions outlined in \cite{andrew_spatial_2026} (e.g., if the data from these channels are saturated at the 4+4 bit quantization level, or is not stationary on $\sim\mathrm{ms}$ timescales). However, \cite{andrew_spatial_2026} has shown that the KLT filter is able to successfully recover transient signals within these two bands using the CHIME telescope. This suggests that the failure to recover these channels using HCO is likely site specific (e.g., incorrect gains, lack of East-West resolution to constrain the direction of the RFI) and work is currently underway to investigate whether further improvements can be made to recover information in these channels. Until then, however, we opt to flag these channels throughout the remainder of our analysis. \par

Overall, the results of Figure \ref{fig:radiometer}, coupled with the smooth-evolving beamformed SEFD in Figure \ref{fig:sefd_test}, imply that HCO's interferometric performance is broadly consistent with theoretical expectations (Paper I, II, III). We emphasize that this statement applies to the majority of HCO's bandwidth, including channels recovered by our updated calibration scheme (Section \ref{s:calibration}).

\subsection{Single-pulse Baseband Sensitivity}\label{ss:baseband_sensitivity}

Next, we assess HCO's interferometric performance within the context of detecting radio transients. In the following section, we showcase (i) how our improved calibration scheme recovers frequency information using a pulsar event (Section \ref{ss:impact_interp}) and (ii) that the achieved sensitivity across our band is broadly consistent with theoretical expectations (Section \ref{ss:autocorr_test}). 

\subsubsection{Impact of Gaussian Process Interpolation \& RFI Filtering on Baseband Sensitivity}\label{ss:impact_interp}

\begin{figure*}
    \centering
    \includegraphics[width=1.0\linewidth]{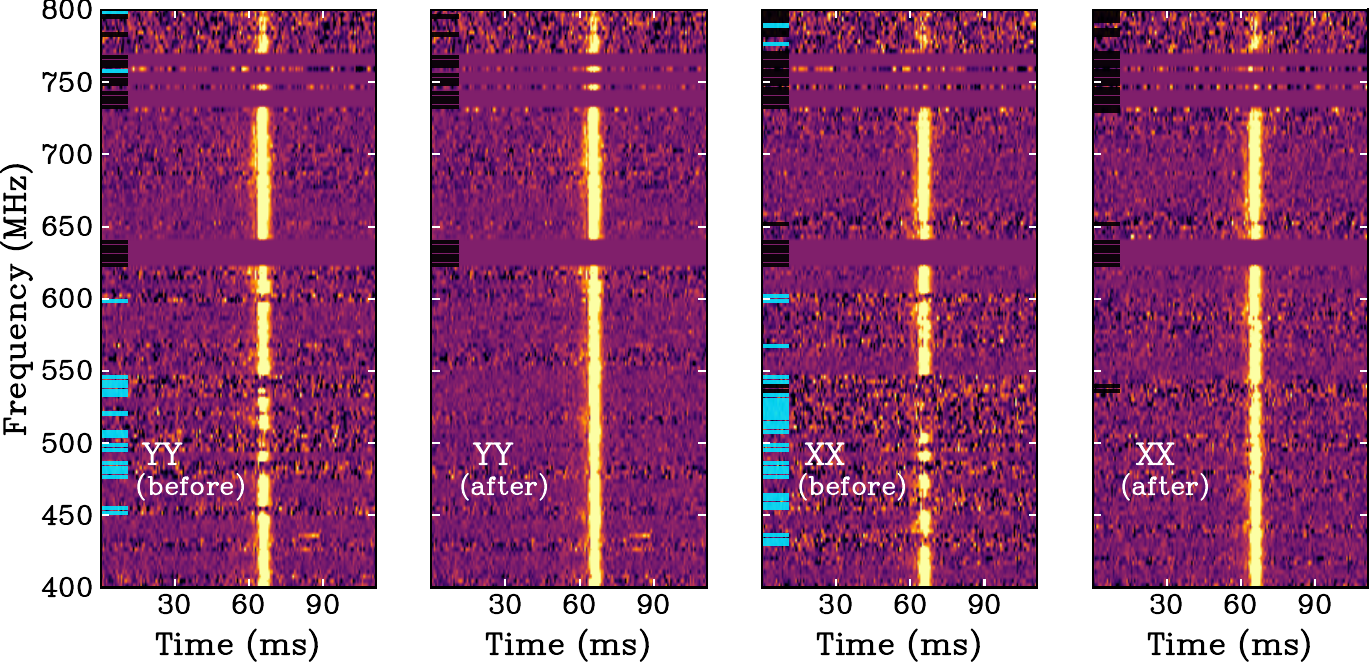}
    \caption{Impact of updating our calibration strategy on our ability to recover astrophysical transient signal in previously RFI contaminated frequency channels. Each panel presents a dynamic spectrum of a PSR B0329+54 pulse, binned in time in $0.66~\mathrm{ms}$ chunks for visual clarity and in frequency in $3.125~\mathrm{MHz}$ chunks. Subplots are grouped by polarization and before/after applying our updated calibration strategy.  Binned frequencies with $\mathrm{S/N < 3}$, computed following Equation  \ref{eq:snr_auto}, are indicated as black and cyan horizontal ticks, representing channels with no significant signal. Cyan ticks represent channels recovered after updating our calibration scheme. \textit{Before:} Updated feed positions, KLT filter and original calibration solutions are applied to the baseband data. \textit{After:} Updated feed positions, KLT filter and updated calibration solutions are applied to the baseband data, isolating improvements of our calibration scheme. }
    \label{fig:before_after_interp_b0329}
\end{figure*}

We use single pulses from PSR B0329+54, a bright ($\gtrsim$ kJy, \citealt{Manchester_2005}) pulsar within HCO's FoV, to measure the performance improvement enabled by the changes described in Section \ref{s:calibration} and \ref{s:feed_positions}. We obtained a full-array, $120~\mathrm{ms}$ baseband capture of a PSR B0329+54 pulse on MJD 61074, originally detected by the CHIME/FRB backend \citep{collaboration_chime_2018} and subsequently triggered at HCO following the process described in Section \ref{ss:triggering}. The corresponding baseband data containing the pulse was saved to disk and transferred to \textsc{canfar} for offline processing. A corresponding dataset captured at CHIME were also saved and transferred to \textsc{canfar} for offline processing. We note that such triggers are routinely captured on an approximately daily cadence at all Outrigger sites and used to diagnose the overall health of the Outrigger systems. \par

For every data product considered in this section, we apply the KLT filter and updated feed positions to solely isolate impovements introduced by out updated calibration scheme. Then, for each version of our calibration solutions (before/after interpolation), we beamform HCO to the known position of PSR B0329+54 accounting for the proper motion of the pulsar \citep{kumar2025}, maximizing the sensitivity of the array in the direction of the transient. Lastly, we coherently dedisperse the baseband data using the known dispersion measure for B0329+54 of $\mathrm{DM} = 26.76$ \citep{Manchester_2005,kumar2025}, and compute the power per polarization. 

To estimate the S/N of the detection as a function of frequency, we compute the autocorrelation S/N for each polarization following: 

\begin{align}\label{eq:snr_auto}
  &\mathrm{S/N}_\mathrm{auto}(\nu) = \notag \\
  &\frac{|\langle\mathcal{V}^*_{\mathrm{H, on}}(\nu) \mathcal{V}_{\mathrm{H, on}}(\nu)\rangle| - \mathrm{med}_g\left[|\langle\mathcal{V}^*_{\mathrm{H, off}}(\nu) \mathcal{V}_{\mathrm{H, off}}(\nu)\rangle_g|\right]}{\mathrm{MAD}_g\left[|\langle\mathcal{V}^*_{\mathrm{H, off}}(\nu) \mathcal{V}_{\mathrm{H, off}}(\nu)\rangle_g|\right]},
\end{align}
where $\mathcal{V}$ is the measured voltage in each polarization as a function of frequency, the ``H" subscript is short-hand for HCO, $\langle\cdot\rangle$ denotes a time-average, $*$ denotes a conjugate and ``on" represents the duration over which the PSR B0329+54 pulse is active. To estimate the noise, the median and MAD are computed over independent ``off" gates which contain no signal from PSR B0329+54, denoted by the $g$ subscript, which are of equal duration to the ``on" gate. Finally, we merge adjacent frequency channels in groups of eight, as this binning is typically done in the fringe search step of the VLBI pipeline.\par

We determine which channels have signal by requiring that $\mathrm{S/N}_\mathrm{auto,binned}\geq3$. We compare the results before and after updating our calibration strategy in Figure \ref{fig:before_after_interp_b0329}, with grey marks highlighting which channels \textit{lack} a significant detection (i.e., $\mathrm{S/N}_\mathrm{auto,binned}<3$).\par

Adopting our original observing strategy, we measure that $71\%$ ($56\%$) of channels have measured signal above the specified threshold for the $Y$ ($X$) polarization. Applying our updated observing strategy, this improves to $86\%$ ($78\%$) for the $Y$ ($X$) polarization, corresponding to a net relative increase of usable bandwidth of $21\%$ ($39\%$). The effect of our updated calibration scheme is most clearly observed at frequencies $\lesssim620~\mathrm{MHz}$, where nearly all of the band is recovered. We note that the lack of detections at frequencies $\gtrsim770~\mathrm{MHz}$ is most likely a combination of gradual degradation of sensitivity at the edges of the bandwidth, alongside the intrinsic source brightness decreasing at higher frequencies, rather than a failure of our calibration scheme. \par

In addition to recovering a significant fraction of our bandwidth, we further observe an \textit{overall} increase in per-channel S/N, with median (excluding RFI contaminated channels) binned S/N increasing from $11.5\rightarrow 17.2$ ($Y$ polarization, $\sim50\%$) and $7.0 \rightarrow 12.2$ ($X$ polarization, $\sim74\%$). This is broadly consistent with the results of our beamformed SEFD measurements in Section \ref{ss:sefd} which found similar instantaneous median sensitivity improvements on the order of $60\%$ ($80\%$) for the $Y$ ($X$) polarization. \par

Overall, the results of Figure \ref{fig:before_after_interp_b0329} provide a concrete example of how our updated calibration scheme successfully recovers astrophysical transient information over channels previously discarded due to RFI contamination. The results further emphasize the importance of valid gain solutions on the success of the KLT filter. Specifically, in the ``before" panel when contaminated solutions are applied to the baseband data, the filter fails to improve the sensitivity in these channels, as expected \citep{andrew_spatial_2026}. However, when the filter is applied in conjunction with valid calibration solutions, we find that HCO's sensitivity is maximized.  In the following section, we quantify this improvement in greater detail. 

\subsubsection{Autocorrelation S/N Test}\label{ss:autocorr_test}

While the results in Section \ref{ss:impact_interp} showcase an improvement in overall sensitivity, in the following section we determine whether this improvement results in a performance that is in-line with theoretical expectations. To test HCO's interferometric performance on transients, we follow section 3.5.2 in Paper II. Briefly, the expected autocorrelation S/N of a transient detected by HCO that is unpolarized with flux density $S_\nu$, duration $\tau$ and bandwidth $\Delta\nu$, after beamforming the array to maximize sensitivity in the direction of the transient, is 

\begin{figure*}
    \centering
    \includegraphics[width=1.0\linewidth]{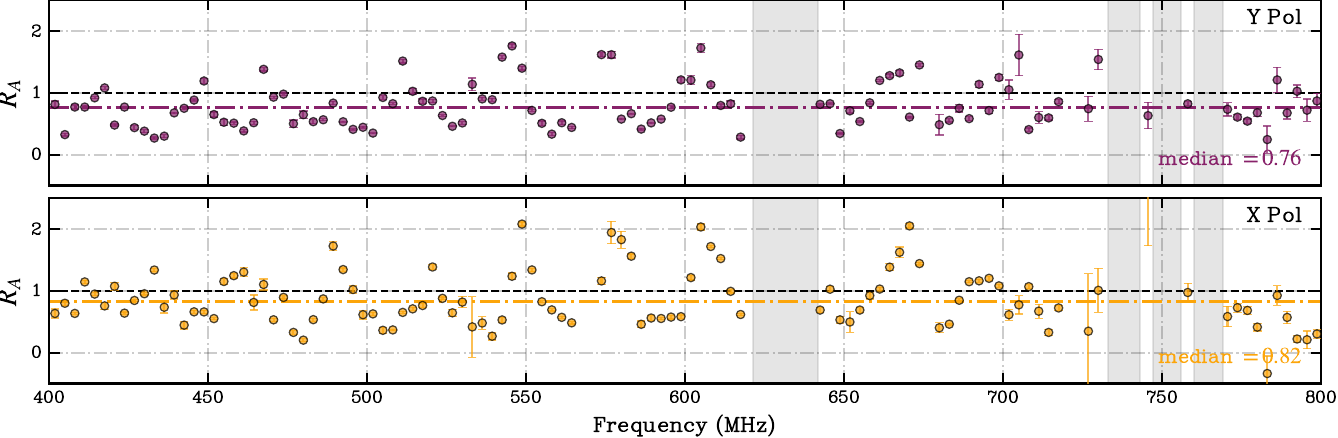}
    \caption{Results of the autocorrelation S/N test (Equation \ref{eq:expected_RA}) using the ``after" PSR B0329+54 pulse plotted in Figure \ref{fig:before_after_interp_b0329}. Measurements of $R_A$  are color-coded based on polarization, and the theoretical expectation of $R_A = 1$ is indicated as a horizontal, black dashed line. The median over non-RFI contaminated channels is indicated in the bottom right of each panel and colored, dash-dotted line. Regions removed due to residual RFI contamination are highlighted by grey shaded regions. Uncertainties are included, but are underestimated with respect to the underlying structure driven by reflections in the cylinders of CHIME and HCO that result in a $30$-MHz standing wave across the band.}
    \label{fig:autocorr_test}
\end{figure*}

\begin{align}
    \mathrm{S/N}_\mathrm{auto} &= \frac{S_\nu}{\mathrm{SEFD_\mathrm{BF}}}\sqrt{\tau\Delta\nu} \\
    &= \frac{S_\nu}{\mathrm{SEFD_\mathrm{inp}}}\sqrt{N(N-1)}\sqrt{\tau\Delta\nu},
\end{align}
where $N$ is the number of independent array elements measuring the arriving signal, and $\mathrm{SEFD}_\mathrm{BF}$ and $\mathrm{SEFD}_\mathrm{inp}$ are the baseband and per-input SEFD measurements described in \ref{ss:sefd}, respectively. However, since the flux density of the source is not known \textit{a priori}, and the SEFD is expected to vary as a function of beam position, we make the simplifying assumption that HCO is approximately identical to CHIME in terms of its overall analog performance. Under this assumption, $\mathrm{SEFD}_\mathrm{inp}$ of both arrays are assumed to be approximately identical. As a consequence, the \textit{ratio} of S/N autocorrelation measurements is expected to scale according to
\begin{align}
    R_\mathrm{A,expected} &= \frac{(\mathrm{S/N}_\mathrm{auto,HCO})}{(\mathrm{S/N}_\mathrm{auto,CHIME})} \\ 
    &= \sqrt{\frac{N_\mathrm{HCO}(N_\mathrm{HCO}-1)}{N_\mathrm{CHIME}(N_\mathrm{CHIME}-1)}}.\label{eq:expected_RA}
\end{align}
Assuming identical performance and full-array participation (e.g., $N_\mathrm{CHIME} = 2048$ and $N_\mathrm{HCO} = 256$), this implies an expected ratio of $R_{A\mathrm{,expected}} \approx 1/8$. \par

To test whether this ratio holds, we perform the test using the same $120~\mathrm{ms}$ baseband capture of PSR B0329+54 shown in Figure \ref{fig:before_after_interp_b0329}. While not shown, an identical capture at CHIME was also acquired. Here, we consider only the \textit{after} pulse (e.g., with improved calibration solutions, feed positions and KLT filter applied to the baseband data) as this represents our current observing strategy. We compute the autocorrelation S/N at both HCO and CHIME following the steps described in Section \ref{ss:impact_interp}. To ensure that the ``on"-pulse and ``off"-pulse gates are identical at both CHIME and HCO, we align the burst arrival time across all frequency channels using \textsc{PyFX} \citep{leung2024vlbisoftwarecorrelatorfast}. We then compute their ratio to measure $R_{A\mathrm{,measured}}$, again merging channels in groups of eight. Defining $R_A \equiv R_{A,\mathrm{measured}}/R_{A,\mathrm{expected}}$, we plot the results in Figure \ref{fig:autocorr_test}. Here, we account for the fact that, when computing $R_{A,\mathrm{expected}}$, both CHIME and HCO have a subset of inputs not actively powered (225 and 51 inputs, respectively) due to known hardware failures. \par

Overall, we measure a median across our band of $R_A$ of $0.76~(0.82)$  for the $Y$ ($X$) polarization. When the KLT filter is \textit{not} applied, the medians degrade by $11\%$ ($5\%$), suggesting that the KLT filter does result in overall improved sensitivity. The clear structure in the ratio is expected and is driven by the $30~\mathrm{MHz}$ standing wave present across both telescopes. Importantly, the results suggest that the performance across recovered frequency channels is in-line with surrounding uncontaminated channels, further validating our updated calibration strategy for this particular event. \par

With both median values falling below 1, we interpret this as potential evidence that $\mathrm{SEFD}_\mathrm{inp}$ (or equivalently, the individual system temperature of HCO inputs) are systematically larger than those of CHIME. However, given the number of simplifying assumptions that go into Equation \ref{eq:expected_RA}, in addition to our results remaining broadly consistent with similar measurements performed by KKO and GBO (Paper II and III), we conclude that HCO is performing within expectation across uncontaminated and recovered frequency channels alike.

\section{Impact of Updated Calibration Scheme on VLBI Performance}\label{s:vlbi_performance}

In the following section, we perform a final set of tests which assess HCO's performance in cross-correlation with CHIME. Specifically, we test whether the measured S/N in cross-correlation with CHIME is within theoretical expectation (Section \ref{ss:xcorr_test}), the improvement in S/N extends to a variety of transient pulse profiles (Section \ref{ss:xcorr_pulsars}), and  the improvement in S/N extends to continuum calibrators (Section \ref{ss:xcorr_continuum}). In doing so, we determine HCO's current capabilities and limitations within the context of VLBI localizing FRBs. 

\subsection{Single-pulse cross-correlation test}\label{ss:xcorr_test}

In Section \ref{s:performance}, we establish the baseline performance of HCO as a standalone interferometer. Here, we determine whether the improved sensitivity in autocorrelation translates to improved sensitivity in cross-correlation with CHIME. This would improve our localization precision for previously-detected bursts and increases the overall number of FRBs that can be localized on the CHIME-HCO baseline \citep{collaboration_chimefrb_2025}. \par

\begin{figure*}
    \centering
    \includegraphics[width=1.0\linewidth]{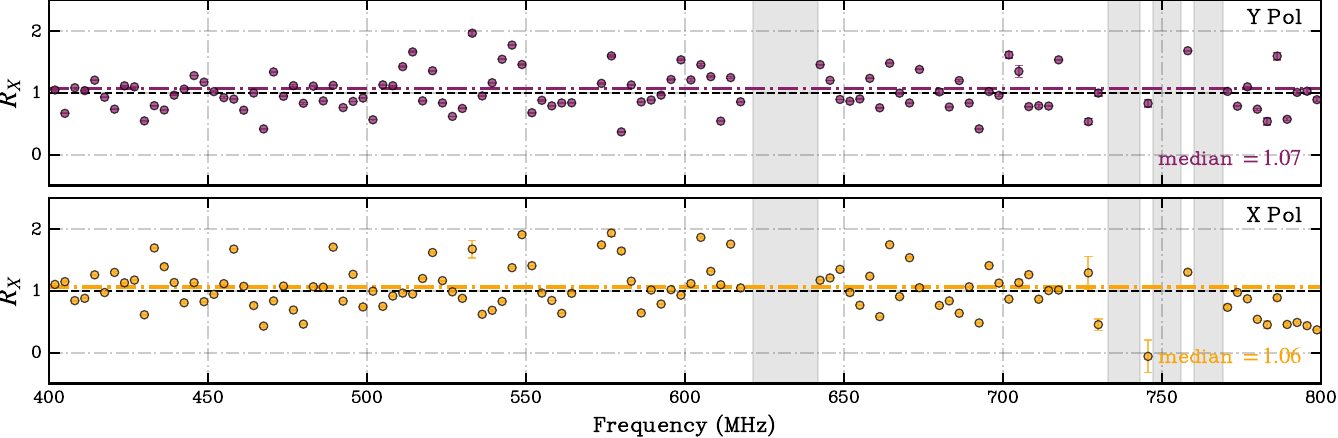}
    \caption{Results of the cross-correlation S/N test (Equation \ref{eq:snr_cross_expectation}) using the ``after" B0329+54 pulse plotted in Figure \ref{fig:before_after_interp_b0329}. $R_X$ measurements (binned by a factor of 8 in frequency to reduce noise) are color coded based on polarization, and the theoretical expectation of $R_X = 1$ is indicated as a horizontal, black dashed line. The median over non-rfi contaminated channels is indicated in the bottom right of each panel and highlighted as a dash-dotted, colored line. Uncertainties are included, but are underestimated with respect to the underlying structure driven by reflections in the cylinders of CHIME and HCO. Regions excluded due to persistent RFI (post filter) are indicated by the shaded grey regions. }
    \label{fig:xcorr_test}
\end{figure*}

To test the performance in cross-correlation with CHIME, we follow Section 4.2 in Paper II. Using the same PSR B0329+54 pulse presented in Section \ref{ss:impact_interp}, we compute the \textit{incoherent} S/N in cross-correlation with CHIME, per polarization, following 
\begin{align}\label{eq:snr_cross}
  &\mathrm{S/N}_\times(\nu) = \notag \\
  &\frac{|\langle\mathcal{V}_{\mathrm{C, on}}(\nu)^* \mathcal{V}_{\mathrm{H, on}}(\nu)\rangle| - \mathrm{med}_g\Big[|\langle\mathcal{V}_{\mathrm{C, off}}(\nu)^* \mathcal{V}_{\mathrm{H, off}}(\nu)\rangle_g|\Big]}{\mathrm{MAD}_g\Big[|\langle\mathcal{V}_{\mathrm{C, off}}(\nu)^* \mathcal{V}_{\mathrm{H, off}}(\nu)\rangle_g|\Big]}.
\end{align}
This is the same as Equation \ref{eq:snr_auto} except that each conjugate product has one factor with subscript C instead of H to denote the cross-correlation between voltage data from HCO and CHIME. Note that the KLT filter is applied to both HCO and CHIME baseband data prior to correlation. The correlation of the signals is performed using \textsc{PyFX} which accounts for and removes known sources of geometric delay. This ensures that the pulse at both sites is aligned to a timing resolution better than the native time resolution of the baseband data ($\leq 2.56~\upmu\mathrm{s}$), necessary for a robust incoherent $\mathrm{S/N}_\times$ measurement. To remove any dependence on the intrinsic properties of the burst (e.g., the flux density), we normalize $\mathrm{S/N_\times}$ by the autocorrelation S/N at CHIME (defined by Equation \ref{eq:snr_auto}, replacing H with C) and again merge adjacent frequency channels in groups of eight. The corresponding ratio is expected to scale according to 
\citep{2015A&C....12..181M, Mena_Parra_2022}
\begin{equation}\label{eq:snr_cross_expectation}
    R_{X,\mathrm{expected}} \equiv \frac{\mathrm{S/N}_\times}{\mathrm{S/N}_\mathrm{auto,C}} = \sqrt{2 \frac{N_\mathrm{H}}{N_\mathrm{C}}},
\end{equation}
where $N$ represents the number of active feeds per station and the factor of $\sqrt2$ accounts for differences in noise statistics between auto- and cross-correlations. Defining $R_X \equiv R_{X,\mathrm{measured}} / R_{X,\mathrm{expected}}$, we compute $R_{X,\mathrm{measured}}$ and normalize by the expected ratio, $R_{X,\mathrm{expected}}$, accounting for feeds removed due to hardware failures. We plot the results in Figure \ref{fig:xcorr_test}. \par

We measure a median across non-flagged channels of $1.07$ ($1.06$) across $Y$ ($X$) polarization, broadly consistent with theoretical expectations. The  $30~\mathrm{MHz}$ structure in the results is again present and expected based on the reflections introduced by the cylindrical design of both arrays. More important than the absolute scaling of the result which is sensitive to simplifying assumptions, we find no clear difference in performance between channels recovered by our updated calibration scheme and those that are uncontaminated. A clear degradation in performance is observed at the upper end of the band in the $X$ polarization. This is consistent with our remarks made in Section \ref{ss:impact_interp} and likely reflects degraded performance at the edge of the band in combination with the faint source brightness expected based on the pulse's spectral behavior between $400$ and $800~\mathrm{MHz}$ \citep{Manchester_2005}. \par

We conclude that HCO's performance in cross-correlation with CHIME is broadly consistent with expectations. This confirms that our improvements in autocorrelation (Section \ref{s:performance}) translate to improved performance in cross-correlation for this particular event. 

\subsection{Cross-correlation sensitivity improvement: pulsars}\label{ss:xcorr_pulsars}

Next, we extend our analysis to a large sample of known pulsars. In doing so, we determine whether the improvements highlighted thus far using a single event translate to consistent performance improvements during day-to-day operations. Specifically, we aim to test whether the improvements continue to persist across bursts that are faint, narrowband or exhibit complex morphology. This contrasts with our tests thus far using only PSR B0329+54 which is one of the brightest pulse-emitting sources observable with HCO. While replicating the tests performed in Sections \ref{ss:autocorr_test} and \ref{ss:xcorr_test} offers the most robust statistic of the system's overall performance on a given day, extending these tests to a broad sample of pulsars is non-trivial. This is due to the tests requiring a sufficiently bright, narrow, broad-band pulse to robustly test the performance across the observing bandwidth. However, many of the pulsars in our sample do not obey these assumptions \citep{Manchester_2005}.\par 

For these reasons, we opt for a simpler metric to assess the overall performance as a VLBI transient detector: 
\begin{equation}\label{eq:integrated_snr}
    \mathrm{S/N}_\times = \sum_\nu \mathrm{S/N}_\times (\nu),
\end{equation}
where $\mathrm{S/N}_\times(\nu)$ is integrated only over channels with confirmed signal. This is done by constructing a dynamic pulse gate both as a function of frequency and time which accurately masks the pulse profile using CHIME voltage data \citep{2021AJ....161...81L,andrew_astrometry_nodate}.  Additionally, we modify Equation \ref{eq:snr_cross} by computing the median and MAD over off-lag frames to estimate the properties of the noise in cross-correlation, rather than specifying temporal off-gates \citep{2021AJ....161...81L,leung2024vlbisoftwarecorrelatorfast}.\par

To test the performance of HCO in cross-correlation, we construct a sample of $59$ events from 14 unique pulsars. These data were obtained using the CHIME/FRB backend \citep{collaboration_chime_2018}, with baseband data captured between MJD 61161 and MJD 61191. Importantly, the month-long separation between events allows us to test the robustness of our calibration scheme over similar time periods. \par

\begin{figure}
    \centering
    \includegraphics[width=1.0\linewidth]{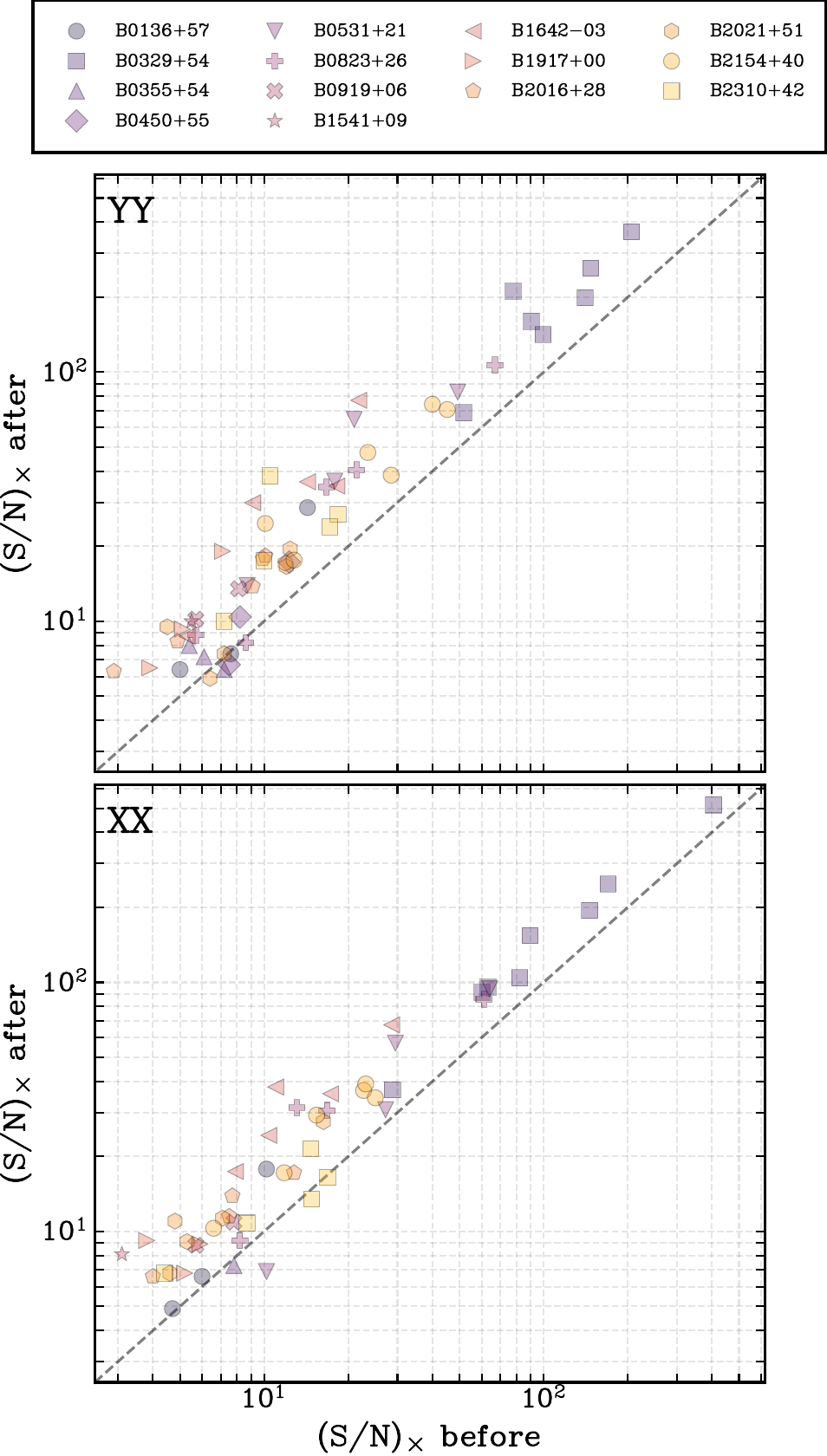}
    \caption{ Comparison between the integrated incoherent cross-correlation S/N of pulsars (Equation \ref{eq:integrated_snr}) before and after updating our observing strategy. \textit{Before:} Default feed positions and default calibration solutions are applied to HCO's baseband data. \textit{After:} Updated feed positions and updated calibration solutions are applied to HCO's baseband data. In either scenario, we do \textit{not} apply the KLT filter to any baseband data. Results are separated by polarization. The sample includes $59$ pulses ($Y$ polarization) and $51$ pulses ($X$ polarization) from $14$ unique pulsars. The dashed diagonal line represents no change in performance, while points above and below represent positive and negative improvement, respectively. Pulsars are identified by their unique color and marker. }
    \label{fig:before_after_xcorr_pulsars}
\end{figure}

For each event, we follow a near-identical procedure outlined in Section \ref{ss:xcorr_test} to correlate the events, beamforming to the known position of each target. The only difference is that we opt to \textit{not} apply the KLT filter across any events in order to isolate improvements driven by our improved calibration scheme and updated feed positions. At the end of this section, we investigate the impact of including the KLT filter. For each event, we compute Equation \ref{eq:integrated_snr} and compare the results before and after applying out updated calibration scheme and feed positions described in Section \ref{s:calibration} and Section \ref{s:feed_positions}. \par

A summary of our results is presented in Figure \ref{fig:before_after_xcorr_pulsars}. Due to the fact that the sensitivity in the $X$ polarization is, on average, worse than the sensitivity in the $Y$ polarization (Section \ref{ss:sefd}), $8$ events failed to correlate in the $X$ polarization both before and after updating our calibration strategy. As such, our final sample consists of $59$ ($51$) events across $Y$ ($X$) polarizations.  We find that, over the majority of pulsar events, we observe an increase in $\mathrm{S/N}_\times$ after applying our updated calibration scheme and feed positions. Importantly, the results apply over a broad range of pulse brightnesses, with $3\leq\mathrm{S/N}_{\times,\mathrm{before}}\leq200$. We measure a median fractional improvement in $\mathrm{S/N}_\times$ of $67\%$ ($52\%$) over the $Y~(X)$ polarization. While our updated calibration scheme drives the \textit{overall} performance improvement, the sample further showcases the impact of our updated feed position determination scheme described in Section \ref{s:feed_positions}. For example, PSR B1642$-$03 is located at the extremities of HCO's beam where the impact of incorrect feed positions is most extreme. Correcting this resulted in an instantaneous improvement in $\mathrm{S/N_\times}$ for this pulsar ranging between $64\%$ ($\mathrm{S/N_\times}$ of $5.3 \rightarrow8.7$) to $241\%$ ($\mathrm{S/N_\times}$ of $21.8 \rightarrow 77.1$) over 5 events. \par

We found that only $5$ out of $59$ events performed worse after applying our changes in the $Y$ polarization and $4$ out of $51$ in the $X$ polarization. However, whenever one polarization performed worse, equal or greater gains were observed in the opposing polarization hand, recovering the losses. We observed no instances in which both polarizations performed worse across all pulsar events. The reason for this behavior is not entirely understood, but suggest that it could be the result of flux moving between the two visibility eigenmodes. When losses did occur, the relative changes were minimal (e.g., at the $1-5\%$ level) and typically associated with intrinsically faint sources ($\mathrm{S/N_\times} \leq 10$). In this regime, we expect some results to degrade based solely on statistical fluctuations alone. Consequently, we conclude that our updated calibration scheme and feed position determination results in consistent sensitivity improvements in cross-correlation of transients on the CHIME-HCO baseline. \par

Finally, we test to what extent the KLT filter described in \cite{andrew_spatial_2026} improves the overall sensitivity on transients in cross-correlation. Re-running the same analysis but applying the filter to both CHIME and HCO data prior to beamforming (in addition to updating the calibration solutions at HCO), we re-correlate the data and compute $(\mathrm{S/N})_\times$ per event. Over the same events presented in Figure \ref{fig:before_after_xcorr_pulsars}, we measure a median increase in S/N of $\sim13\%$ across both polarizations. Note that this increase is relative to the \textit{after} data in Figure \ref{fig:before_after_xcorr_pulsars} (e.g., post updating our calibration strategy). We do find, however, that the KLT filter does result in some events performing worse in both polarizations. However, this behavior occurs in less than $\lesssim10\%$ of all events, suggesting that the majority of events benefit from employing the KLT filter. While further investigation into site-specific requirements to optimize the performance of the KLT filter is underway, the above results suggest that the combination of improved calibration solutions and KLT filter represents the current best observational setup for maximizing the number of transients localizable by HCO. \par

We note that by considering only the incoherent cross-correlated S/N, we are blind to the impact of our updated calibration scheme on the \textit{astrometric} performance of the array. However, preliminary results strongly suggest that our updated calibration strategy does \textit{not} introduce a bias in the astrometry of observed sources at the level required for precision localization of FRBs. This analysis is beyond the scope of this paper and will be presented in detail in upcoming works \citep{andrew_astrometry_nodate}. 

\begin{figure}
    \centering
    \includegraphics[width=1.0\linewidth]{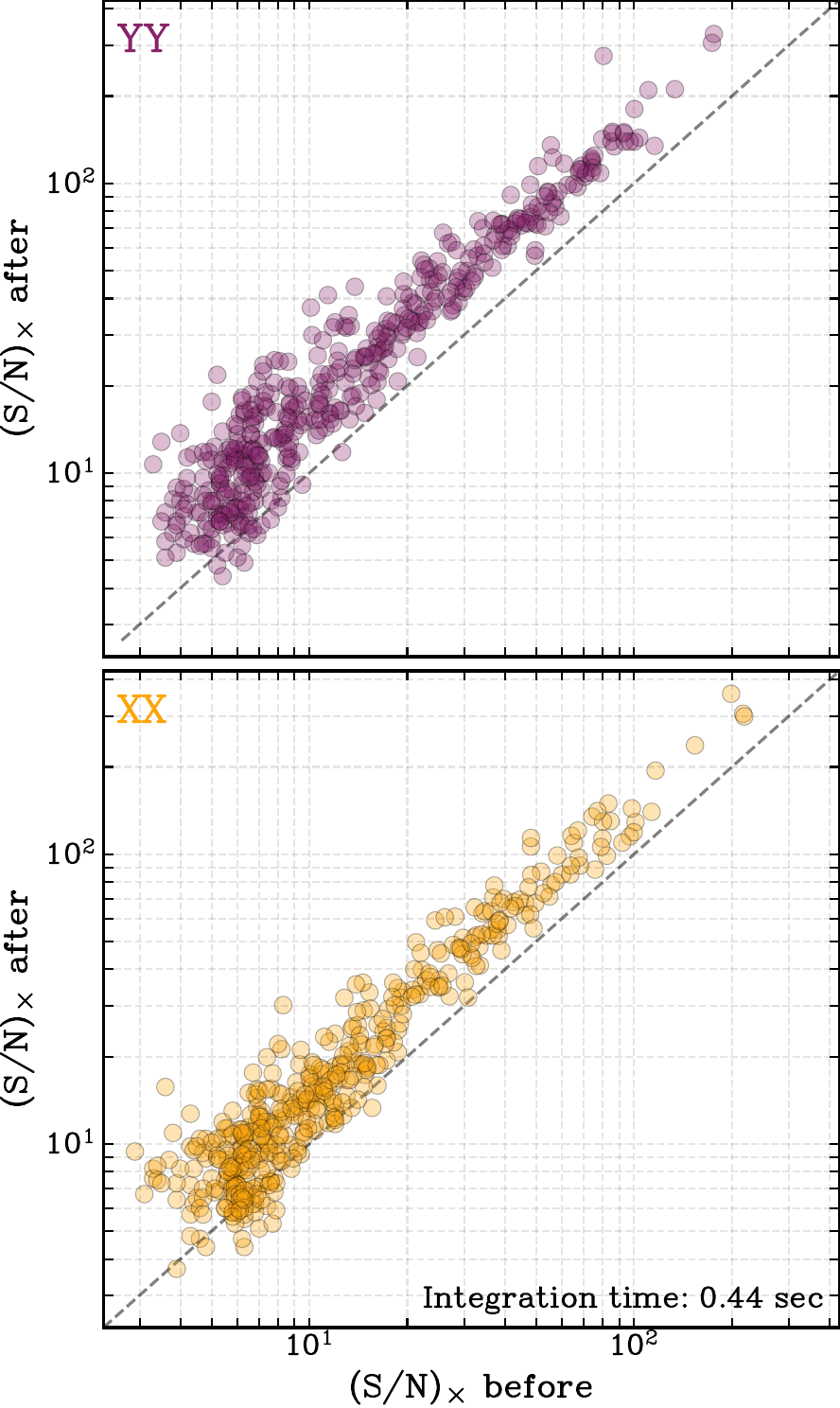}
    \caption{Same as Figure \ref{fig:before_after_xcorr_pulsars}, this time using continuum calibrators integrated over $440\,\mathrm{ms}$. The sample includes 465 visibility pairs from 126 unique calibrators from the Radio Fundamental Catalog \citep{rfc_2025}. }
    \label{fig:xcorr_snr_calibrators_400ms}
\end{figure}

\subsection{Cross-correlation sensitivity improvement: continuum sources}\label{ss:xcorr_continuum}

As a final test, we wish to assess whether the performance increase observed in our pulsar sample further translates to improvement in $(\mathrm{S/N})_\times$ when observing VLBI continuum calibrators \citep{rfc_2025}. Here, continuum calibrators refer to radio sources (often active galactic nuclei) that are bright ($\gtrsim100~\mathrm{mJy}$), whose positions are known to $\lesssim1~\mathrm{mas}$ and that are located in proximity (sky-projected separation $\lesssim10^\circ$) to the FRB to calibrate direction dependent sources of delay. In the context of the CHIME/FRB Outriggers, the ionosphere is often the most dominant source of direction dependent delay on the very long baselines \citep{collaboration_chimefrb_2025,2025ApJ...981...39A,andrew_astrometry_nodate}. However, since the position of an FRB is not known \textit{a priori}, it is often the case that the Outriggers must decide whether to prioritize a more nearby, faint calibrator versus a more distant, bright calibrator \citep{collaboration_chimefrb_2025, andrew_astrometry_nodate}. By improving HCO's overall sensitivity, we increase the number of viable calibrators in HCO's FoV, allowing for further optimization of our VLBI calibration strategy. \par

To test the HCO-CHIME baseline sensitivity on continuum calibrators, we utilize a subset of the sample in Section \ref{ss:xcorr_pulsars} and follow a near identical strategy as when we carried out the test using pulsars. However, rather than beamform in the direction of the transient, we beamform both CHIME and HCO to all known calibrators from the Radio Fundamental Catalog \citep{rfc_2025} within CHIME and HCO's FoV. Note that in this case, we forego applying the KLT filter due to the large sample ($\gtrsim400$ pointings) and computational cost associated with running the filter. Per polarization, we correlate the signals using \textsc{PyFX}, with the second difference being that we average over the length of the baseband capture, rather than the duration of the transient. For these events, the typical duration is around $440~\mathrm{ms}$. Finally, we compute Equation \ref{eq:integrated_snr} and repeat this exercise to compare between our original and updated calibration strategy. \par

Our final sample consists of 465 visibility pairs from 126 unique VLBI calibrators from the Radio Fundamental Catalog \citep{rfc_2025}. We plot the results comparing the performance before and after updating our calibration scheme in Figure \ref{fig:xcorr_snr_calibrators_400ms}. We measure a median increase in $\mathrm{S/N}_\times$ of $64\%$ ($44\%$) in the $Y$ ($X$) polarization. In the low S/N regime, we observe larger scatter in S/N improvements in comparison to the high S/N regime. As a result of these improvements, calibrators that were previously too weak $(\mathrm{S/N}_\times \leq 4)$ are now strong enough to use for calibration ($\mathrm{S/N}_\times \geq 15$). Out of $465$ visibility pairs, only $3\%$($7\%$) were found to have reduced $\mathrm{S/N}_\times$ across $Y$($X$) polarization. Again, however, when sensitivity decreased in one polarization, the sensitivity was entirely recovered by the opposing polarization hand. A similar behavior was also seen in the pulsars observed in Section \ref{ss:xcorr_pulsars}.  \par

Repeating a similar exercise but increasing the integration time to $1.4~\mathrm{s}$ (corresponding to the current hardware limit of the CHIME/FRB Outrigger systems), we find similar improvements. In a sample of 104 visibility pairs from 53 unique sources integrated over $1.4~\mathrm{seconds}$, we measure an equal median improvement of $32\%$ across both polarization hands. While we still measure an overall improvement in performance, the decrease in the median likely suggests that RFI is beginning to contribute more significantly to (but not dominate) the noise budget over longer timescales. This has important implications as the Outriggers team is actively developing ``tracking beams" (i.e., the ability to continuously readout tied-array voltage data at a specified pointing) to increase the integration time of calibrators to tens of seconds~\citep{Pearlman+2024c}. In combination with the improvements highlighted throughout this work, the density and detection significance of calibrators within $\lesssim10^\circ$ of the target is expected to dramatically increase, enabling precise localizations of an even greater number of FRBs. Furthermore, improvements to the VLBI calibrator search presented in \cite{2025ApJ...981...39A}, utilizing our updated calibration scheme and longer integration times, are currently underway will be presented in upcoming works \citep{atkinson2026}.   \par

Based on these results, we conclude that the improvements observed in interferometric performance further translate to improved sensitivity in cross-correlation of continuum calibrators. In comparison to our original calibration strategy, the CHIME-HCO baseline now has access to a larger sample of calibrators across HCO's entire FoV, with particular emphasis on those at large zenith angles.

\section{Summary and Conclusion}\label{s:conclusion}

This work presents the commissioning results of HCO, the third and final Outrigger station in the CHIME/FRB Outrigger array. Motivated by a complex RFI environment that consistently contaminates $\sim40\%$ of HCO's observing bandwidth, we develop and implement an updated calibration strategy to mitigate its impact. This relies on simple and conservative assumptions about the behavior of the complex gain solutions (e.g., that they are smoothly evolving across the band). We then use GPR to recover calibration solutions over RFI-corrupted channels. In doing so, we find that HCO's performance is in line with theoretical expectation for $\gtrsim80\%$ of its observing bandwidth over timescales relevant for transient research ($\lesssim\mathrm{seconds}$).\par

Our results have important implications for precisely localising a large sample of FRBs. First, by improving HCO's beamformed SEFD by $\gtrsim60\%$ (Section~\ref{ss:sefd}) while maintaining thermal noise properties (Section~\ref{ss:radiometer_test}), we increase the sensitivity of the array to fainter bursts by a factor of $\sim60\%$ in cross-correlation  (Section~\ref{ss:xcorr_pulsars}). As a result, we increase the sample of FRBs that are detected with sufficient S/N to be localized by the full array. Second, while the astrometric performance of the array will be described in upcoming works \citep{andrew_astrometry_nodate}, the improvement in sensitivity translates to improved astrometric precision given the increased S/N per event in cross-correlation with CHIME. Third, the improved sensitivity enables access to a wider sample of in-beam calibrators used to localize FRBs. This allows for direction-dependent effects to be more robustly constrained, resulting in improved localization accuracy. Fourth, the sustained improvement over longer timescales has implications for ongoing hardware improvements to enable $10\times$ longer integration times on calibrators. Our results suggest that HCO will continue to benefit from our updated calibration strategy, resulting in dramatic improvements in calibrator density surrounding FRB events. Finally, when our calibration strategy is applied to all Outrigger sites, we expect the overall sensitivity of the array to also improve. Altogether, the results of this work enable HCO and the Outriggers to accelerate their ability to precisely localize FRBs. \par

Within the broader context of large-$N$ arrays, our RFI filtering and gain interpolation methods may be beneficial to several upcoming next-generation interferometric arrays, including CHORD \citep{vanderlinde_lrp_2019}, BURSTT \citep{lin_burstt_2022}, DSA \citep{hallinan+2019_dsa}, CHARTS \citep{CHARTS_design}, CASM \citep{connor_256-antenna_2026} and SKA-LOW \citep{timmerman2026lowfrequencyvlbiskalow}. Importantly, all of these arrays share a common strategy of utilizing sky information to calibrate individual antenna responses across the array. Simultaneously, all of these arrays are also interested in maximizing the number of detected and localized transients. However, each array must grapple with RFI environments that are unique to each individual site, and which vary in their complexity. In the case of HCO, we find that successful implementation of simple algorithms leads to the recovery of $\sim97$ effective inputs worth of data, $\sim37\%$ of the total array. \par

Our work emphasizes that even in an RFI environment that contaminates nearly the majority of the sky-inferred complex solutions, significant sensitivity improvements can be achieved by adopting simple algorithmic techniques offline. We note that some degree of fine-tuning will be required to accommodate array- and site- specific requirements. However, our example adds to the growing body of literature (e.g., \citealt{van_der_veen_signal_2004, Finlay_2023, zhang2024rfidrunetrestoringdynamicspectra, andrew_spatial_2026, kuiper_tied-array_2026}) motivating the development of techniques aimed at recovering astrophysical information in noise and RFI-contaminated data, departing from the standard practice of discarding contaminated data outright. This is of particular importance in the current era where RFI will only continue to occupy larger fractions of allocated bandwidth (e.g., \citealt{Di_Vruno_2023,Bassa_2024}) traditionally used for radio astronomical research.

\section*{Acknowledgments}
Funding for the CHIME/FRB Outrigger program is provided by a grant from the Gordon \& Betty Moore Foundation. CHIME is funded by a grant from the Canada Foundation for Innovation (CFI) 2012 Leading Edge Fund (Project 31170) and by contributions from the provinces of British Columbia, Qu\'{e}bec and Ontario. The CHIME/FRB Project, which enabled development in common with the CHIME/Pulsar instrument, is funded by a grant from the CFI 2015 Innovation Fund (Project 33213) and by contributions from the provinces of British Columbia and Qu\'{e}bec, and by the Dunlap Institute for Astronomy and Astrophysics at the University of Toronto. Additional support was provided by the Canadian Institute for Advanced Research (CIFAR), McGill University and the McGill Space Institute thanks to the Trottier Family Foundation, and the University of British Columbia. The CHIME/Pulsar instrument hardware was funded by NSERC RTI-1 grant EQPEQ 458893-2014. This research was enabled in part by support provided by the BC Digital Research Infrastructure Group and the Digital Research Alliance of Canada (alliancecan.ca).

M.L. acknowledges the support from the Natural Sciences and Engineering Research Council of Canada (NSERC-CGSD). J.M.P. acknowledges the support of an NSERC Discovery Grant (RGPIN-2023-05373). A.P.C. is a Canadian SKA Scientist and is funded by the Government of Canada / est financ\'e par le gouvernement du Canada.  M.D. is supported by a CRC Chair, NSERC Discovery Grant, and CIFAR. V.M.K. holds the Lorne Trottier Chair in Astrophysics \& Cosmology, a Distinguished James McGill Professorship, and receives support from an NSERC Discovery grant (RGPIN 228738-13). C. L. acknowledges support from the Miller Institute for Basic Research at UC Berkeley. K.W.M. is supported by NSF Grant No. 2510771. A.B.P.~acknowledges support by NASA through the NASA Hubble Fellowship grant \mbox{HST-HF2-51584.001-A} awarded by the Space Telescope Science Institute, which is operated by the Association of Universities for Research in Astronomy, Inc., under NASA contract \mbox{NAS5-26555}. A.B.P.~also acknowledges prior support from a Banting Fellowship, a McGill Space Institute~(MSI) Fellowship, and a Fonds de Recherche du \mbox{Qu\'ebec -- Nature} et Technologies~(FRQNT) Postdoctoral Fellowship. V.S. is supported by a Fonds de Recherche du Quebec---Nature et Technologies (FRQNT) Doctoral Research Award. The AstroFlash research group at McGill University, University of Amsterdam, ASTRON, and JIVE is supported by: a Canada Excellence Research Chair in Transient Astrophysics (CERC-2022-00009); an Advanced Grant from the European Research Council (ERC) under the European Union's Horizon 2020 research and innovation programme (`EuroFlash'; Grant agreement No. 101098079); an NWO-Vici grant (`AstroFlash'; VI.C.192.045); an NSERC Discovery Grant (RGPIN-2025-06681); an ERC Starting Grant (`EnviroFlash'; Grant agreement No. 101223057); and an NWO-Veni grant (VI.Veni.222.295).

The authors acknowledge the use of the Canadian Advanced Network for Astronomy Research (\textsc{canfar}) Science Platform operated by the Canadian Astronomy Data Centre (CADC) and the Digital Research Alliance of Canada, with support from the National Research Council of Canada (NRC), the Canadian Space Agency (CSA), CANARIE, and the Canada Foundation for Innovation (CFI).

\software{PyFX \citep{leung2024vlbisoftwarecorrelatorfast},
          kotekan \citep{andre_renard_2021_5842660},
          scikit-learn \citep{scikit-learn},
          matplotlib \citep{Hunter:2007},
          emcee \citep{2013PASP..125..306F},
          numpy \citep{harris2020array}
          }
          
\bibliography{main}{}
\bibliographystyle{aasjournal}

\end{document}